\documentclass[sigconf,authorversion,nonacm]{acmart}
\AtBeginDocument{%
  \providecommand\BibTeX{{%
    \normalfont B\kern-0.5em{\scshape i\kern-0.25em b}\kern-0.8em\TeX}}}

\usepackage{multirow}
\usepackage{colortbl}
\usepackage{graphicx,scalerel}
\usepackage{boldline}

\begin{document}

\title[I Can Embrace and Avoid Vagueness Myself]{I Can Embrace and Avoid Vagueness Myself: Supporting the Design Process by Balancing Vagueness through Text-to-Image Generative AI}

\author{Myungjin Kim}
\affiliation{%
  \institution{Hanyang University}
  \city{Seoul}
  \country{Republic of Korea}}
\email{myungkim331@hanyang.ac.kr}

\author{Bogoan Kim}
\affiliation{%
  \institution{Hanyang University}
  \city{Seoul}
  \country{Republic of Korea}}
\email{bogoankim@hanyang.ac.kr}

\author{Kyungsik Han}
\authornote{Corresponding author}
\affiliation{%
  \institution{Hanyang University}
  \city{Seoul}
  \country{Republic of Korea}}
\email{kyungsikhan@hanyang.ac.kr}

\renewcommand{\shortauthors}{Kim., et al.}

\begin{abstract}
This study examines the role of vagueness in the design process and its strategic management for the effective human-AI interaction. While vagueness in the generation of design ideas promotes diverse interpretations and prevents fixation, excessive vagueness can lead to scattered results. Designers attempt to use image search tools or generative AIs (e.g., Dall-E) for their work but often fail to achieve satisfactory results because the level of vagueness is not properly managed in these technologies. In this work, we identified how designers coordinate vagueness in their design process and applied key components of the process to the design of CLAY, an interactive system that balances vagueness through iterative prompt refinement by integrating the strengths of text-to-image generative AI. Results from our user study with 10 fashion designers showed that CLAY effectively supported their design process, reducing design time, and expanding creative possibilities compared to their existing practice, by allowing them to both embrace and avoid vagueness as needed. Our study highlights the importance of identifying key characteristics of the target user and domain, and exploring ways to incorporate them into the design of an AI-based interactive tool.
\end{abstract}



\begin{teaserfigure}
  \includegraphics[width=\textwidth]{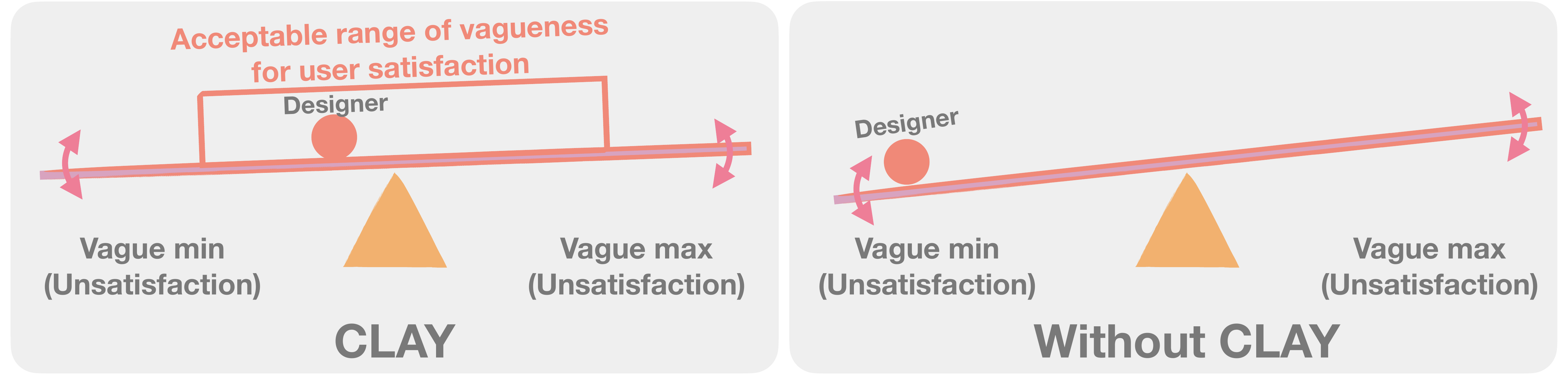}
  \centering
  \caption{The diagram illustrates how CLAY (left) helps designers maintain an optimal balance of vagueness for user satisfaction, preventing the process from becoming too vague or too precise. Without CLAY (right), designers can easily exceed this acceptable range, leading to dissatisfaction.}
  \Description{The image compares two diagrams illustrating the concept of vagueness in design. On the left, labeled "CLAY," a balanced seesaw represents the "Acceptable range of vagueness for user satisfaction." A designer figure sits at the center, maintaining balance between "Vague min" and "Vague max," both labeled as causing unsatisfaction. On the right, labeled "Without CLAY," the seesaw is unbalanced, tilting towards "Vague max." This shows how CLAY helps designers maintain an optimal balance of vagueness, while without it, designers can easily exceed the acceptable range, leading to user dissatisfaction.}
  \label{fig:teaser}
\end{teaserfigure}


\maketitle

\section{Introduction}
For designers, the notion of vagueness in design plays various roles. Designers’ initial design ideas often begin as uncertain and abstract mental images~\cite{stacey1999sketch, eckert2000sources, howell2016value}. This vagueness allows for multiple interpretations of the design, such as providing a rich array of design strategies, fostering the transformation of ideas, and preventing early fixation that limits creativity to pre-existing concepts~\cite{minneman1991social, fish1990amplifying, tseng2018can, goel1995sketches}. However, when vagueness becomes too excessive, it can lead to scattered or directionless results due to the lack of clear constraints in the design development process~\cite{stacey2003against}. Therefore, effective management of vagueness---within its appropriate range, which tends to vary from person to person---is crucial to the success of a project~\cite{steinmann1997modellbildung, leite2011analysis, howell2016value}. In this paper, we refer to this process as balancing vagueness.

Vagueness facilitates the exploration of meaningful information~\cite{tseng2018can, wu2022ai}, allowing designers to explore a wide range of alternatives and gradually reduce vagueness as the designers uncover more specific information~\cite{kuhlthau2004seeking, chowdhury2011uncertainty}. Image search tools (e.g., Google Image Search, Pinterest) are commonly used for this process by providing diverse search results that aid in the exploration of relevant information~\cite{linder2014everyday}. However, even for specific queries, these tools often produce inaccurate or irrelevant results, complicating the search process and increasing the time required to find meaningful information~\cite{spink2001searching, woll2005user, liu2022opal}. This can lead to an excessive increase in vagueness, which complicates design tasks and hinders effective decision-making~\cite{whittemore1973generalized}. On the other hand, the emergence of text-to-image generative AI offers the potential to quickly generate meaningful and high-quality images through prompts, bypassing the need for multiple search steps in the design process~\cite{liu2022opal}. However, achieving meaningful results requires specific, well-designed prompts, which may be beyond the capabilities of designers and limit the creative advantages designers can gain from exploring various possibilities when embracing vagueness~\cite{ko2023large}. Therefore, it is important to explore the best ways to leverage the advantages of text-to-image generative AI within the vagueness balancing process.

Previous research has explored the use of text-to-image generative AI to support the design process by refining prompts to better capture users' creative intent~\cite{wang2024promptcharm, brade2023promptify, mahdavi2024ai}. However, these studies have largely overlooked the creative design aspect of exploiting or reducing vagueness. It is also important to balance the level of vagueness, avoiding making it too large or too small, otherwise, users are likely to stop using the technology. With these motivations, this study addresses the gap between the unique characteristics of design processes, the requirements of technology use, and user experience (expectation, satisfaction, tolerance) in two important ways: (1) by leveraging the advantages of text-to-image generative AI to mitigate the excessive vagueness that is often encountered with traditional image search tools, and (2) by helping designers gain valuable design strategies and support the vagueness balancing process. While advances in generative AI show great potential for creative work, its application remains limited because users often struggle to align their mental models with the prompts required by these AI models~\cite{wu2022ai, zamfirescu2023johnny}. The HCI community has emphasized the importance of designing generative AI systems that align with users' existing mental models~\cite{weisz2024design}. Thus, our study seeks to contribute to the field of HCI by exploring how integrating text-to-image generative AI with the mental model of the vagueness balancing process can more effectively support the creative process.

In this paper, we present CLAY, an interactive system designed to support the process of vagueness balancing through iterative prompt refinement, metaphorically likened to molding undefined ideas like clay into a complete design. Given the importance of vagueness in the fashion domain~\cite{stacey1999sketch}, the design rationale of CLAY was developed through a formative study with six senior fashion professionals, all of whom had experience with text-to-image generative AI. We conducted interviews with the participants focusing on (1) how designers move from initial vagueness to a final design when creating a moodboard; (2) the exploration process designers undertake to gather meaningful information from image search tools while balancing vagueness, and the challenges they face; and (3) the difficulties they encounter when using text-to-image generative AI in creative processes. Based on the interview results, we identified three key insights: (1) the role of hierarchical and combinational structures in balancing vagueness, (2) the challenges of navigating both intentional and unintentional interactions during the vagueness balancing process, and (3) the difficulties of designing prompts for text-to-image generative AI in the context of vagueness balancing. From these insights, we derived three design goals for CLAY: (1) support designers in balancing vagueness through hierarchical and combinational structures; (2) facilitate the transition from hierarchical, unintended outcomes to combinational, intended ones; and (3) support prompting by maintaining vague prompts and helping to specify them to better balance vagueness.

CLAY operates with the following main structure: the user starts by creating a vague prompt, explores the hierarchical results generated by the text-to-image generative AI to take advantages of vagueness, refines the prompt, and receives combination results from the AI to reduce vagueness. The main structure is divided into two sub-structures: (1) When the user enters a vague prompt related to the initial concept, CLAY generates sub-styles and sub-fashion elements in text format. The user refines the prompt, and CLAY creates a moodboard. (2) When the user enters a moodboard, CLAY suggests fashion elements that can be used in the design. The user then refines the prompt based on these suggestions, and CLAY generates corresponding designs. Through this iterative process, CLAY aims to support the user's design workflow.

We conducted a within-subjects study with 10 fashion professionals to compare CLAY with the baseline system (Dall-E). The results showed that (1) the steps in CLAY allowed users to control the results according to their intentions, leading to a reduction in the number of interactions; (2) CLAY helped expand the scope of the moodboard and even allowed users to modify its components, enabling them to create the desired moodboard more proactively; and (3) users applied the generated images more actively in their final designs, exploring by adding keywords that were not suggested by the system, and entering more specific keywords to create the desired images more actively. Based on these results, we discussed the importance and role of text-to-image generative AI in improving the clarity of vagueness. Specifically, (1) understanding the vagueness explore structure to let users actively embrace vagueness; (2) understanding the appropriate timing and prompt guidance to let users actively avoid vagueness; (3) providing users to select the system based on vagueness level; and (4) providing the real images in materializing the design.

In summary, this paper makes the following contributions:
\begin{itemize}
    \item We identified the hierarchical combinational structure in the vagueness balancing process, the challenges posed by traditional image search tools and input prompting, and how these issues can be addressed using text-to-image generative AI.
    \item We developed CLAY, a system that efficiently supports the design process by incorporating the vagueness balancing process into text-to-image generative AI.
    \item Our process confirmed that users can embrace or avoid vagueness as needed, and we discussed the possible directions and design implications for text-to-image generative AI based on these findings.
\end{itemize}

Our study contributes to expanding the current research on creative support tools (CST) by demonstrating how text-to-image generative AI can be applied to creative processes that require balancing vagueness. We suggest that future systems designed with these characteristic processes in mind can be more effectively applied to similar domains. Identifying key elements (e.g., vagueness in a fashion domain in our case) that can be considered important for characterizing domain users, tasks, or the domain itself, and extending the concept to design elements for the interactive computer system would be important for making the system more meaningful and adaptable to domain users.

\section{Related Work}
\subsection{Vagueness in the Creative Design Process}

In this paper, we focus specifically on vagueness, which is related to a lack of preciseness in the design process~\cite{klir1987we, dictionary1989oxford}. Vagueness occurs when ideas are unclear and difficult to distinguish, making it challenging to determine a direction. Many designers form rough ideas about shape, layout, and size in their minds~\cite{stacey1999sketch}. At this stage, rather than making concrete decisions, these ideas remain in a state where the specific details are not fully defined~\cite{stacey1999sketch}. For example, rough shapes and compositions may appear as sketchy lines~\cite{suwa2000unexpected}.

Fashion design is a domain where this vagueness is common. Stacey et al.~\cite{stacey1999sketch} specifically discuss vagueness in the context of knitwear designers. Knitwear designers often think of creating 'one of those,' where they have a specific idea in mind, but this idea only serves as a temporary element to represent a broader, more abstract category. Even when they envision the exact garment they want to create, there is still a lack of detailed information, and the design continues to evolve as they further refine the details. In other words, designers may envision a clear form, but the detailed elements remain undefined, making vagueness an inevitable part of the fashion design process.

\subsection{Negative and positive impact of vagueness in the design process}

People generally experience vagueness as an unpleasant state and are motivated to reduce or eliminate it~\cite{bar2009feeling}. The greater the uncertainty, the more difficult the decision may be. If not enough effort is not invested in refining the vagueness, it can lead to serious problems. For example, design biases may occur, costly maintenance may be required, and confusion about the design direction may arise~\cite{stacey2003against, rajabalinejad2012incorporating, morse2018tolerancing}. Therefore, the act of reducing vagueness in the design process has been emphasized as important~\cite{rajabalinejad2012incorporating}.

However, it should be noted that vagueness can play an important role in the creative process~\cite{morse2018tolerancing}. Designers use their vagueness to ensure that all possible interpretations are considered. This leads to the generation of various design alternatives and the transformation of creative ideas~\cite{tversky2011creativity}. It also helps to avoid early fixation (i.e., limiting oneself to previously used concepts~\cite{jansson1991design}) during the design development process~\cite{fish1990amplifying, tseng2018can, goel1995sketches, schon2017reflective, purcell1998drawings}. For example, when sketching, designers intentionally visualize the vagueness of the current state to foster creative ideas~\cite{stacey1999sketch, tseng2018can}, which promotes creativity and innovation~\cite{fish1990amplifying, tseng2018can, scrivener1993interaction}. This creative transformation becomes the driving force behind design innovation~\cite{tseng2018can, liu1996designing}. Ultimately, vagueness is useful for reframing design problems and plays an important role in finding more creative and innovative solutions through deeper reflection.

Thus, vagueness in the design process stimulates creativity and encourages the exploration of multiple possibilities. As decisions are made incrementally over time, design solutions become more refined and detailed~\cite{aakerman2012contextualist, abualdenien2020vagueness, tracey2016uncertainty}. The decisions made during this phase can have a critical impact on the success of a project~\cite{leite2011analysis, howell2016value}. Maximizing the benefits of vagueness while compensating for its drawbacks is critical, and in this paper, we refer to this process as "balancing vagueness."

\subsection{Gathering information to balance vagueness}
The process of exploring various possibilities of vagueness while gradually reducing it to search for specific information (balancing vagueness) unfolds over time~\cite{kuhlthau2004seeking, chowdhury2011uncertainty}. When faced with vagueness, an internal conflict often arises~\cite{mishel1981measurement, mishel1991uncertainty}, and designers seek the necessary information to resolve this conflict~\cite{edelman1997curiosity, berlyne1970novelty}. When information is lacking, it negatively affects the decision-making process and outcomes by increasing the vagueness of how the design will evolve due to the many undecided factors~\cite{knotten2015design}. Thus, the process of finding meaningful information from vagueness is essential~\cite{tseng2018can, wu2022ai}.

In the design domain, computer-based design tools have supported this process~\cite{li2022analyzing, palani2021active, zhang2019understanding, yamaguchi1992probabilistic}. Specifically, image search tools (e.g., Google Image Search, Pinterest) are useful for exploring design variations by allowing users to view a wide range of images based on search terms~\cite{linder2014everyday}. Nearly 60\% of queries are formulated with only one or two terms, making the information retrieval process more complex~\cite{spink2001searching}. Furthermore, on these platforms, users are often overwhelmed by a great number of irrelevant results~\cite{liu2022opal}. When vagueness increases too much due to irrelevant information, the value of being in a decision state decreases~\cite{whittemore1973generalized}. Therefore, while existing image search tools offer the benefit of generating diverse ideas, they can sometimes make it difficult to balance vagueness due to an overabundance of information.

On the other hand, the emergence of text-to-image generative AI offers the advantage of generating infinite images that match the user's intent without going through a complex process, compared to traditional image search tools~\cite{liu2022opal}. To generate images that match the intended results, more precise prompts are required. However, users have expressed a desire to explore different variations of vagueness, and they mentioned that if they have to enter specific prompts from the beginning, they are unable to explore ideas from vagueness~\cite{ko2023large}.

Therefore, we aim to explore ways to support prompt composition using text-to-image generative AI, which can effectively leverage both the positive and negative aspects of vagueness, and address the limitations of traditional search processes. In doing so, we aim to help users effectively balance vagueness while working on their desired tasks.

\subsection{Human and text-to-image generative AI collaboration in vagueness}

Ko et al.~\cite{ko2023large} emphasized the need for prompt support to assist visual artists in their creative work with text-to-image generative AI. Research has been presented that uses prompt refinement to enable the exploration of various designs and the generation of accurate images. Liu et al.~\cite{liu20233dall} developed a system that supports prompt creation by recommending relevant components, styles, and designs as users input their design goals. This system prevented design fixation among CAD designers and helped generate new design ideas. Brade et al.~\cite{brade2023promptify} developed a system to support designers' creativity through random images. PromptMagician by Feng et al.~\cite{feng2023promptmagician} improved final image results by fine-tuning input prompts and hyperparameters through multi-level visualization. Hao et al.~\cite{hao2024optimizing} proposed a method that fine-tunes a pre-trained language model on a small dataset and uses reinforcement learning to generate aesthetically superior images while preserving the user's intent. These previous studies demonstrate that by refining prompts, designers can explore various ideas and generate highly accurate images. However, there is a lack of research that focuses on vagueness balancing, which allows designers to remain in the initial concept-based vagueness in order to maximize the diversity of the output while simultaneously mitigating the decision-making difficulties that can result from excessive vagueness. There is also limited research on how this feature can be incorporated into a system to improve the user experience.

Wang et al.~\cite{wang2024promptcharm} discovered the potential for exploration-
exploitation from PromptCharm, a system that automatically refines and improves prompts to effectively interpret the intentions of novice users. They noted that it is important to consider the balance between exploration-exploitation when designing interactive systems for prompt engineering and creative design. In addition, as Louie et al.~\cite{louie2020novice} pointed out, the human-AI interface for creative design should empower users "whether or not they have a clear creative goal in mind." These suggestions imply the importance of considering the characteristics of the creative process when designing such systems. Based on the need for this research, we propose that considering the balance of vagueness is one of the key factors in maximizing creativity through human-AI collaboration.

In this study, we present a prompt refinement system for text-to-image generative AI that helps to effectively balance vagueness in the creative process, specifically in the domain of fashion design. To achieve this, we conduct a formative study to identify the elements necessary for the balancing vagueness process, evaluate their impact on the user experience, and discuss design implications for future balancing vagueness systems.

\section{Formative Study}

We conducted a formative study with fashion professionals to identify key considerations for AI-human interaction design aimed at balancing vagueness.

\subsection{Participants and recruitment}
We recruited participants based on their ability to take a design from the ideation stage to the final product. Additionally, participants must have used text-to-image generative AI at least once in their design process. A total of six participants were recruited through a web community where fashion professionals share fashion trends, job opportunities, and style tips. The final participants included one male and five females, with fashion design experience ranging from 2 to 4 years, which is sufficient to cover the full design process. This study was approved by the IRB at the authors’ institution. Each participant was compensated \$30 for their participation.

\subsection{Study procedure}
The design process typically consists of two main phases: the creation of a moodboard, which presents the overall concept and direction of the upcoming design using various images, and then drawing inspiration from the moodboard to develop the design itself~\cite{mckelvey2011fashion}. Based on this process, we aksed the participants three main interview questions. First, we asked how designers complete their designs from initial vagueness while creating a moodboard and designing. The goal of this question was to understand the designers' existing process and structure for balancing vagueness. Second, we asked about the process of using image search tools to obtain meaningful information during the vagueness balancing process, and then to describe the challenges they encountered. The goal was to identify the challenges designers face in balancing vagueness when using image search tools, and to propose solutions to these challenges using text-to-image generative AI. Third, we asked about the challenges of using text-to-image generative AI during the creative process. This question focused on identifying difficulties in using the AI during the vagueness balancing process and offering potential solutions.

Two authors of this paper independently coded the interview transcripts and repeatedly discussed the coding results to categorize themes until the coders reached a consensus.

\subsection{Formative study results}
We identified key ideas to support the balancing of vagueness. We mapped each challenge to the corresponding "design goal" as follows.

\begin{figure*}[]
\vspace{-0.3cm}
\centering
\includegraphics[width=0.8\textwidth]{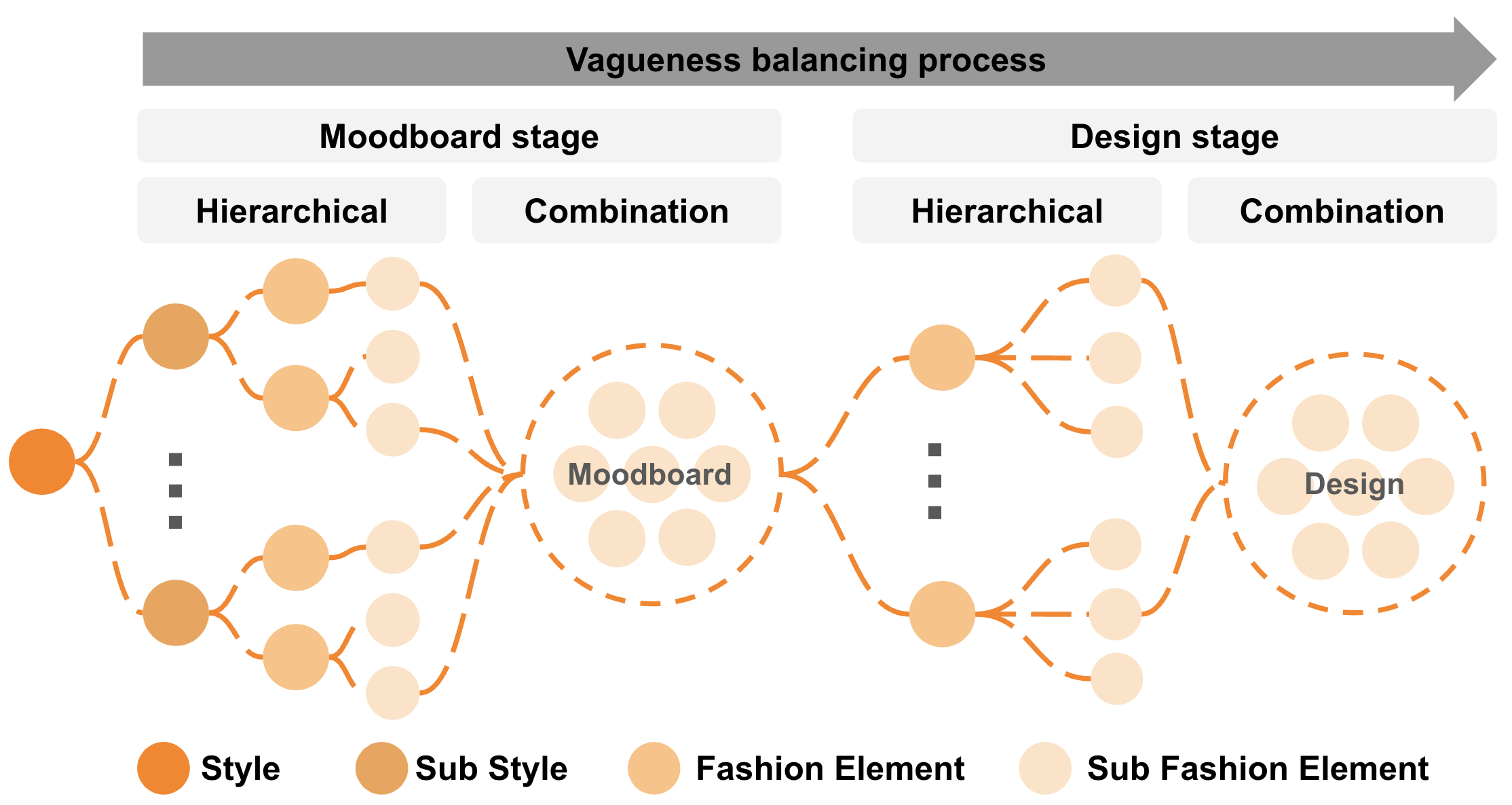}
\vspace{-0.2cm}
\caption{The diagram shows the hierarchical and combinational structures used in the vagueness balancing process for both the moodboard and design stages. For each stage, the process begins with a hierarchical structure, and as users progress through the process, these elements are combined. The number of circles is a placeholder and the variety could be more extensive.}
\Description{The image illustrates the vagueness balancing process for moodboard and design stages. It shows a horizontal flow from left to right, divided into two main sections: Moodboard and Design. Each section contains hierarchical and combination structures represented by interconnected circles. The hierarchical structure starts with a 'Style' (darkest circle), branching out to 'Sub Styles', then to 'Fashion Elements', and finally to 'Sub Fashion Elements' (lightest circles). The combination structure is shown as a cluster of circles labeled 'Moodboard' and 'Design' respectively. The diagram uses a color gradient from dark orange to light peach to represent the hierarchy. Dotted lines connect the hierarchical structures to the combination clusters, illustrating the progression of the vagueness balancing process.}
\label{fig:HC}
\end{figure*}

\subsubsection{Limited understanding of the structures used by fashion designers in balancing vagueness.}
Since little is known about how fashion designers balance vagueness through specific structures, we investigated how designers attempt to deal with vagueness. We found that fashion designers start with vague ideas and explore this vagueness using a hierarchical approach, followed by a combination process to reduce vagueness and finalize their designs. This hierarchical and combination process was evident in both the moodboard creation and design development phases (Figure~\ref{fig:HC}).

\begin{figure*}[]
\vspace{-0.2cm}
\centering
\includegraphics[width=0.9\textwidth]{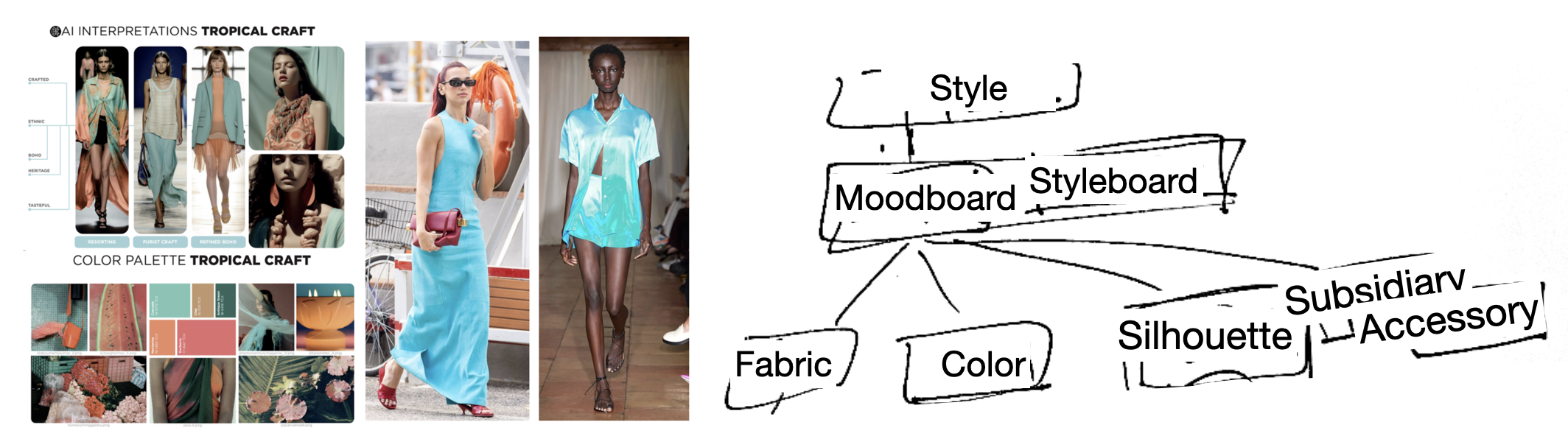}
\vspace{-0.2cm}
\caption{(Left) A moodboard that is typically created by the designer. (Right) A sketch of the moodboard, which is likely to be expanded or modified (with different elements or structures) over time.}
\Description{The image is divided into two parts. On the left is a moodboard titled "TROPICAL CRAFT" featuring fashion photographs and a color palette. The photos show models wearing light, flowy garments in tropical colors like turquoise, orange, and pastel hues. The color palette displays swatches of these tropical colors. On the right is a hand-drawn sketch diagram representing the structure of a moodboard. It shows a hierarchical relationship starting with "Style" at the top, branching down to "Moodboard" and "Styleboard," which further branch into elements like "Fabric," "Color," "Silhouette," and "Subsidiary Accessory." This sketch illustrates how a moodboard can be expanded or modified over time with different elements or structures. CopyRetry}
\label{fig:MD_ex}
\end{figure*}

\textbf{Hierarchy and combination in the moodboard process:} When designers decide on the style they want to pursue in their design, they create a moodboard that allows room for various interpretations of that style and stays within a space of vagueness. \textit{"To create a moodboard like this, you first decide on a style. [...] But the concept of a style is not that simple and can be interpreted in many ways. For example, within the feminine style, there are minimal feminine, romantic feminine, and so on. [...] This is a minimal, feminine moodboard with a summer resort feel that I created (Figure~\ref{fig:MD_ex})"} (P1). After exploring different interpretations, the designer then combines fashion elements to compose the moodboard. \textit{"The elements that make up a moodboard are very diverse. The top and bottom clothing items, accessories, shoes, bags, hairstyles, and even the background and colors are carefully chosen (Figure~\ref{fig:MD_ex})"} (P4).

\begin{figure*}[]
\centering
\includegraphics[width=0.6\textwidth]{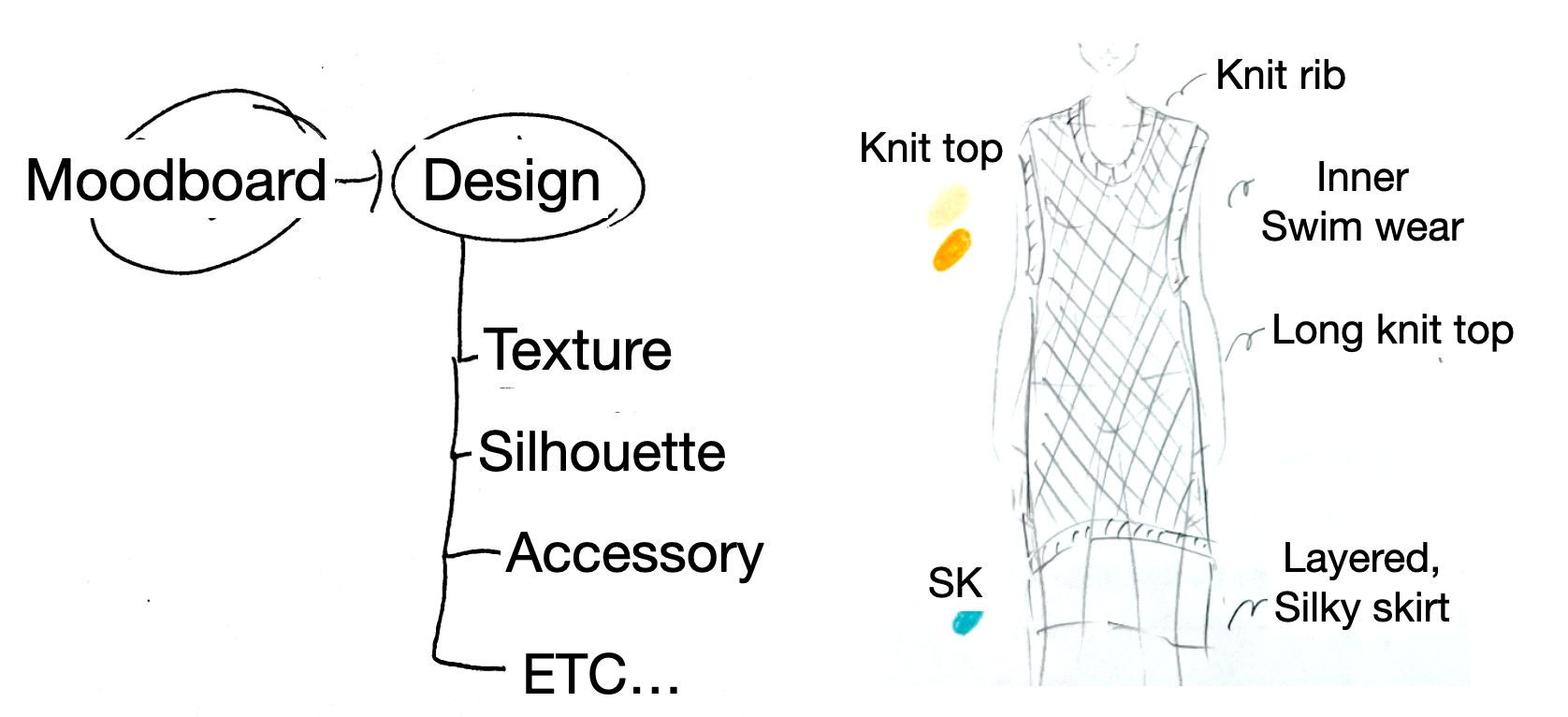}
\vspace{-0.2cm}
\caption{The left image shows a sketch of how designs are derived from the moodboard. Subcomponents in the moodboard are likely to be modified depending on the level of vagueness. The right image shows a sample design based on the conventional method.}
\Description{The image is divided into two parts. On the left is a hand-drawn sketch showing the relationship between "Moodboard" and "Design". The "Design" element branches out into subcomponents: Texture, Silhouette, Accessory, and ETC. This illustrates how designs are derived from the moodboard, with subcomponents potentially modified based on vagueness levels. On the right is a sketched fashion design of a layered outfit. It features a long knit top with a rib detail, worn over an inner swimwear, and paired with a layered, silky skirt. The sketch is annotated with labels pointing out these specific elements, demonstrating a sample design based on the conventional method.}
\label{fig:DG_ex}
\end{figure*}

\textbf{Hierarchy and combination in the design process:} Designers often leave room for multiple interpretations when incorporating moodboard elements into their designs, effectively using vagueness. \textit{"When incorporating elements from the moodboard into the design, I explore how they might be interpreted like this (Figure~\ref{fig:DG_ex}). For example, I consider whether to use a particular element should be used in the texture or as an accessory, and if it's for the texture, what form the element should be transformed"} (P3). After exploring various possibilities, the fashion elements are combined into a single design, reducing the vagueness. \textit{"This is a design I created based on the moodboard. I really liked the color composition from the moodboard, so I used it. The flowing shape of the skirt dress was beautiful, so I applied it to the skirt swim wear, and I thought this top would look nice as knitwear, so I designed it that way (Figure~\ref{fig:DG_ex})"} (P1).

\begin{itemize}
    \item \textbf{DG1. Help designers balance vagueness based on the hierarchical and combination structure results.} We aim to use a hierarchical structure result to take advantage of vagueness, and a combinational structure result to avoid excessive vagueness.
\end{itemize}

\subsubsection{Difficulties in transitioning from unintended to intented results while balancing vagueness.}
\begin{figure*}[]
\centering
\includegraphics[width=0.8\textwidth]{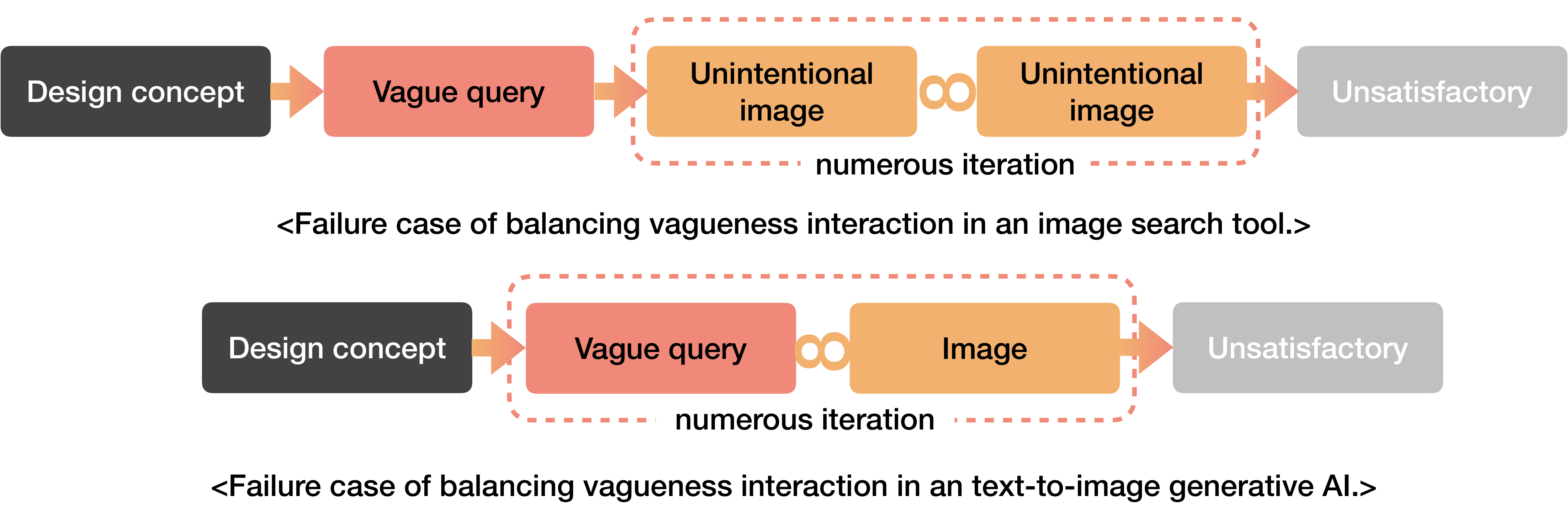}
\vspace{-0.2cm}
\caption{Failure cases for balancing vagueness in both an image search tool (top) and a text-to-image generative AI (bottom). In both scenarios, numerous queries with no chance of getting to the next step lead to high user dissatisfaction and abandonment.}
\Description{The image shows two flow diagrams illustrating failure cases in balancing vagueness for design tools. The top diagram, labeled "Failure case of balancing vagueness interaction in an image search tool," depicts a process from "Design concept" to "Vague query" to multiple "Unintentional image" steps, ending in "Unsatisfactory." The bottom diagram, labeled "Failure case of balancing vagueness interaction in a text-to-image generative AI," shows a similar process but with a single "Image" step instead of multiple unintentional images. Both diagrams highlight "numerous iteration" between the vague query and image generation steps, emphasizing how repeated unsuccessful attempts lead to user dissatisfaction and abandonment in both image search and AI-generated image scenarios.}
\label{fig:PI}
\end{figure*}

We confirmed that designers continue to explore images using search tools (e.g., google image search and pinterest) based on the hierarchical combination structure identified earlier. During this process, designers go through a process of exploring various interpretations of their initial vague ideas by observing unintended results, guided by the hierarchical structure. Then, through the combined structure, they gradually reduce the vagueness and arrive at a more intentional and concrete output. However, they mentioned that reducing vagueness takes a long time (Figure~\ref{fig:PI}). \textit{"When I want to create a feminine moodboard, I type in 'feminine' as a keyword [...] It helps to keep searching through different results is helpful, but if unrelated results keep coming up, it takes too long to complete the moodboard"} (P4). One participant also mentioned that they sometimes stop using image search tools at the final stage because it is difficult to find the images they want. \textit{"I often can't think of how to use an element in my design. So, I just randomly search for the name of the element. [...] But, when I'm finalizing designs, I stop using Pinterest because it’s hard to find images that I want exactly."} (P6).

\begin{itemize}
    \item \textbf{DG2. Support the transition from hierarchical unintended results to combination intended results to balance vagueness.} We aim to support unintended exploration through a hierarchical process that exploits vagueness, while guiding designers to intended results through a combination process that reduces vagueness, which is not possible with existing image search tools.
\end{itemize}

\subsubsection{Difficulties in crafting detailed prompts while balancing vagueness.}
Despite the advantages of text-to-image generative AI in quickly achieving intended outcomes as identified in previous research, designers mentioned that it is difficult to engage more deeply with the tool due to the challenges of prompting, especially because of the initial vagueness  (Figure~\ref{fig:PI}). \textit{"I tried it once, but I didn't like it [...] Well, someone says you have to write in detail, but design isn't something that's fully defined from the start. I'm often not sure what to say during the design process, so writing prompts feels overwhelming"} (P5). They also found it challenging to express their thoughts clearly in the prompts, when reducing vagueness. \textit{"It's hard to describe the image I want in detail. I want to find an image of a certain thing, but I don't know how to articulate it."} (P1).

\begin{itemize}
    \item \textbf{DG3. Support the prompting by maintaining vague prompts and helping to specify prompts to balance vagueness.} We aim to ensure that users do not have to fully specify prompts in the early stages due to initial vagueness, while also helping them with the challenges of crafting detailed prompts throughout the process of reducing vagueness.
\end{itemize}

\section{CLAY}
In this paper, we present CLAY (a metaphor for shaping undefined ideas like clay to complete a design), an interactive system that supports the process of balancing vagueness through iterative prompt refinement using DALL-E and GPT.

\begin{figure*}[]
\centering
\includegraphics[width=\textwidth]{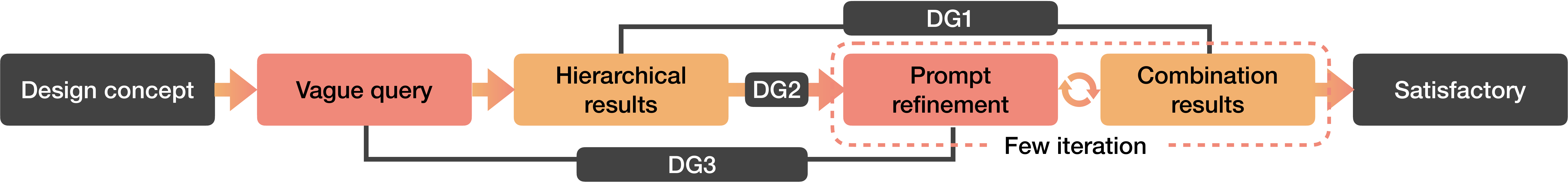}
\caption{The main process structure of CLAY consists of four stages: starting from the design concept, users enter a vague query that leads to hierarchical results. They then refine the prompt and iterate a few times until they reach the desired combination of results, achieving user satisfaction.}
\Description{The image shows a flow diagram of the CLAY process structure. It consists of six stages connected by arrows: "Design concept" leads to "Vague query," then to "Hierarchical results," followed by "Prompt refinement," "Combination results," and finally "Satisfactory." The process includes three decision gates (DG1, DG2, DG3) that allow for iterative refinement. DG1 spans from "Hierarchical results" to "Combination results," DG2 connects "Prompt refinement" and "Combination results" with a "Few iteration" loop, and DG3 links back from "Prompt refinement" to "Vague query." This structure illustrates how users can refine their prompts and iterate through the process to achieve satisfactory results from an initially vague design concept.}
\vspace{-0.2cm}
\label{fig:CI}
\end{figure*}

\subsection{CLAY main functions}
The user starts by composing a "vague prompt," which reduces the pressure to enter detailed prompts (DG3). Next, they explore the hierarchical keyword results generated by the text-to-image generative AI, using the vagueness to explore various possibilities (DG1). The user then refines the prompt using the keywords provided by CLAY, avoiding being overwhelmed by unintended hierarchical results (DG2), and receiving assistance in specifying prompts (DG3). Next, CLAY generated the results from the keyword user input, which helps reduce vagueness (DG1). This framework supports the balancing of vagueness across four stages (Figure~\ref{fig:CI}). These phases are applied to both the moodboard and the design process.

\subsubsection{Balancing vagueness in the moodboard process}
The user begins by entering a prompt based on keywords related to a style and mood. CLAY then generates a hierarchy of sub-styles, fashion elements, and sub-elements derived from the style and mood keywords in a text format. The user explores the results and reviews the suggested keywords. Next, the user refines the prompt using the keywords suggested by CLAY. Finally, the user can view the moodboard created based on the keywords entered, which helps to reduce vagueness. This process is shown in Figure~\ref{fig:CIM}.

\subsubsection{Balancing vagueness in design process}
\begin{figure*}[]
\centering
\includegraphics[width=0.9\textwidth]{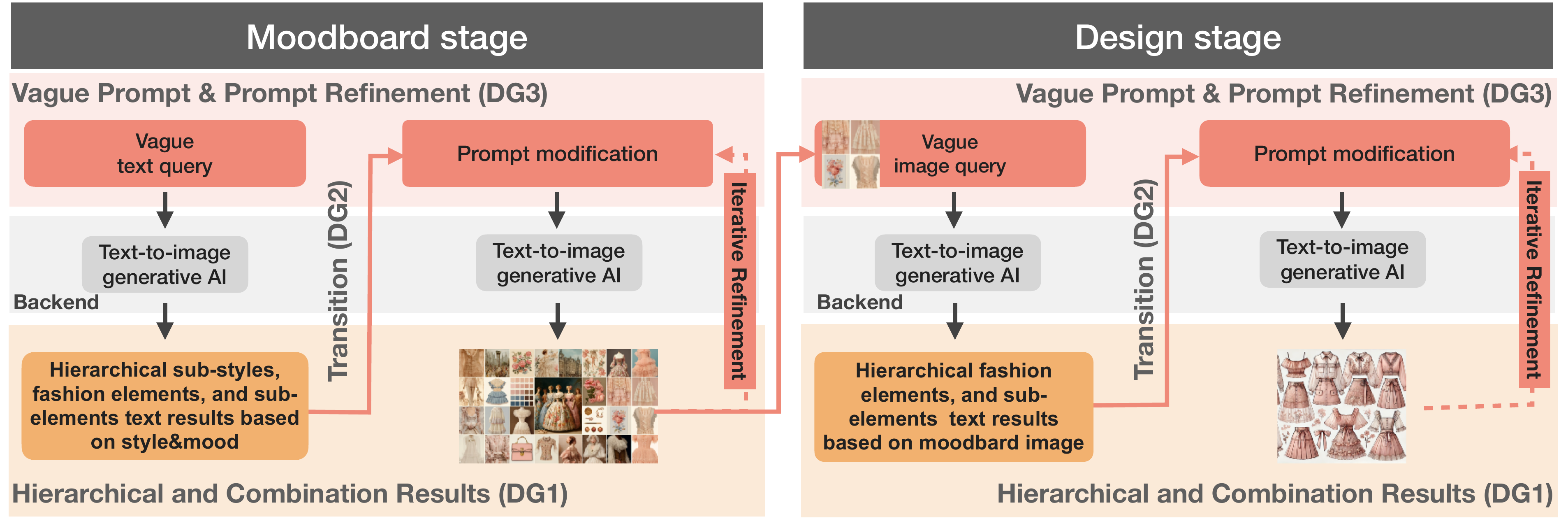}
\caption{The flow of CLAY's four main structures as applied to the moodboard and design process. The diagram demonstrates the progression from vague prompts to hierarchical results, which transitions into prompt refinement and combination results for both the moodboard and design stages.}
\Description{The image illustrates CLAY's four main structures applied to the moodboard and design process, divided into two parallel columns. The left column shows the Moodboard stage, and the right shows the Design stage. Both stages follow a similar flow: starting with a vague prompt (text query for moodboard, image query for design), moving through a text-to-image generative AI backend, producing hierarchical results, and then entering a prompt modification and refinement loop. The moodboard stage generates hierarchical sub-styles and fashion elements based on style and mood, while the design stage produces fashion elements based on the moodboard image. Both stages show example image outputs: a collage of vintage-style clothing for the moodboard, and a set of similar dress designs for the design stage. The diagram emphasizes the iterative nature of the process, with transitions (DG2) and iterative refinement loops clearly marked.}
\vspace{-0.2cm}
\label{fig:CIM}
\end{figure*}

When the user inputs a moodboard, the system provides various hierarchical suggestions in the form of fashion element and sub-element keywords that can be used in the design. The user explores these keywords, maximizing the advantages of vagueness by considering the multiple design possibilities that can be interpreted from the moodboard. Based on the keywords suggested by CLAY, the user refines the prompts to gather more specific information. The system then generates a design by combining the input fashion sub-elements, allowing the user to reduce vagueness. This process is illustrated in Figure~\ref{fig:CIM}.

\subsection{Technical description}
When transitioning from a vague prompt to hierarchical results, the prompt suggestion engine is implemented using the GPT-3.5 (ChatGPT) and DALL-E models, both developed by OpenAI~\cite{openai_api}. For the combination results, we used DALL-E's text-to-image generation~\cite{ramesh2021zero}.

We used zero-shot learning, except for keyword extraction. In zero-shot learning, no concrete examples other than instructions are provided to the model. Since the LLM is highly effective at generating diverse subject suggestions with only the given instruction~\cite{brade2023promptify}, adding more examples could introduce bias and reduce the exploration space, which would be counterproductive to the balancing vagueness required by our process.

When the user first enters keywords related to style and mood, the captioning results are processed to extract lists of styles and moods. We used few-shot prompting, which uses a small set of task-specific examples to guide the LLM in generating output that follows similar patterns. Next, based on the extracted keywords for style and mood, we use zero-shot prompting to generate a hierarchy of sub-styles, fashion elements, and sub-element keywords. After the user refines the prompt with more detailed input, we again extract lists of fashion elements, and use DALL-E to generate moodboard images. When the user enters a moodboard image, we instruct the LLM to generate image captions that describe the fashion elements that can be drawn from the moodboard. After the user further refines the prompt, we again extract lists of fashion elements, DALL-E is used to generate different designs.

\subsection{User scenario}

\begin{figure*}[]
\vspace{-0.3cm}
\centering
\includegraphics[width=1.0\textwidth]{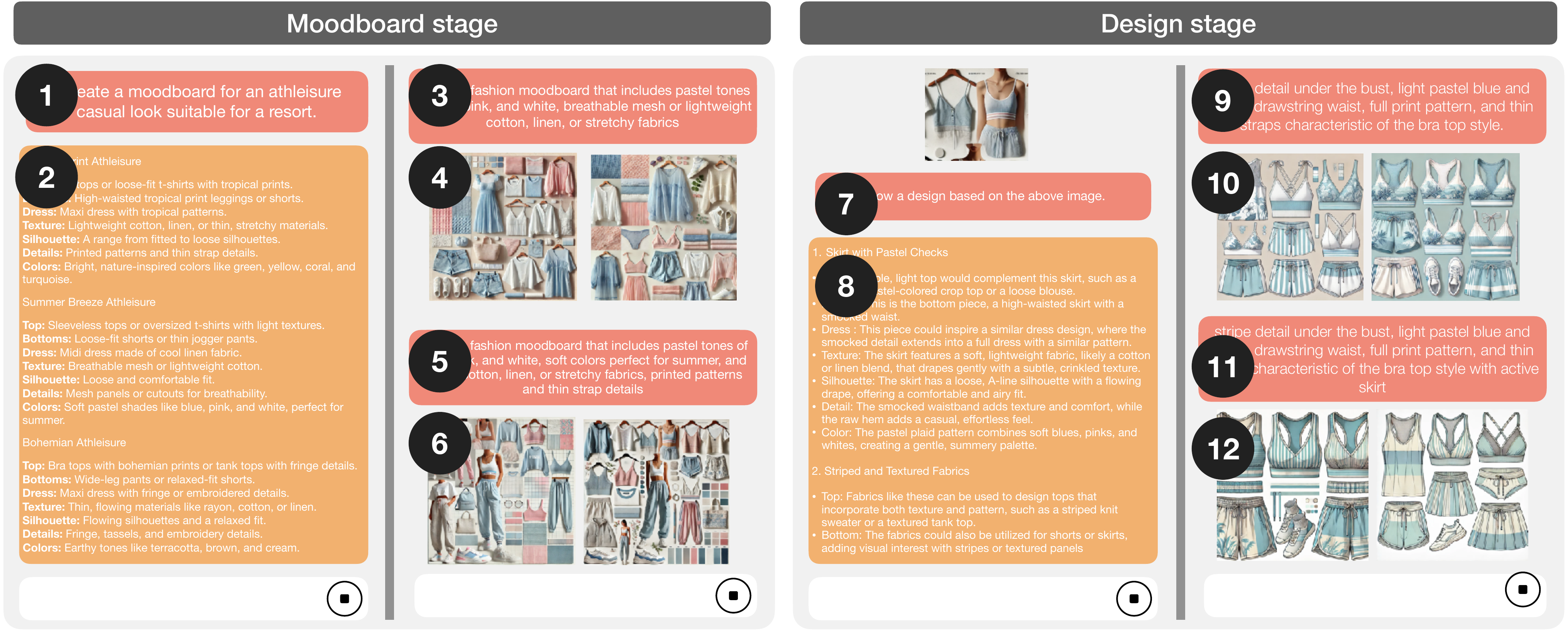}
\vspace{-0.2cm}
\caption{This figure illustrates the "Moodboard stage" on the left and the "Design stage" on the right. In the moodboard stage, (1) participants enter a design concept, (2) the system then presents hierarchical results. (3-6) The user next refines the prompt and views the moodboard combination results through multiple iterations. (7) The design process follows a similar approach, starting with a moodboard-based design concept and (8) viewing hierarchical results. (9-12) The user then refines the prompt and views design combination results through multiple iterations.}
\Description{This image illustrates two stages in a design process: the "Moodboard stage" on the left and the "Design stage" on the right. In the Moodboard stage, participants are prompted to create a moodboard for an athleisure or casual look suitable for a resort. Several examples are presented, such as "Tropical Athleisure," "Summer Breeze Athleisure," and "Bohemian Athleisure," each describing various outfit elements like tops, bottoms, dresses, materials, silhouettes, and details. The system then generates results based on prompts, such as incorporating pastel tones of cotton, linen, or stretchy fabrics. Participants can refine the prompt through multiple iterations while viewing hierarchical results of the moodboard combinations. In the Design stage, participants create a design based on an image from the moodboard. For example, details such as pastel blue, light fabrics, and specific design features like drawstrings or stripes are highlighted. The design process is also iterative, allowing users to refine the prompt and view hierarchical design combinations to arrive at the final outcome.}
\label{fig:CIF}
\end{figure*}

Scenario-based approaches in Human-Computer Interaction (HCI) research are effective because they help articulate the use situation from multiple perspectives~\cite{carrol1999five}. By illustrating how users might engage with CLAY and balance vagueness, we offer a user-centered perspective that emphasizes the goals that users may adopt and pursue.

To illustrate the use scenario of CLAY, we introduce Jenny, a senior fashion design student, who uses CLAY to create a design for a fashion competition. Jenny decided to participate in a fashion design contest organized by a fashion magazine with the theme "Athleisure Casual Look." Since the athleisure casual look was trend this year, many innovative designs had already been created, so Jenny needed a design that would stand out. Drawing on her experience from last semester, where she worked on a resort theme, Jenny thought she could focus on athleisure casual wear appropriate for a resort setting. However, the combination of unique keywords made it difficult to find images that could help define her design direction, despite using various image search tools. Faced with this challenge, Jenny decided to use CLAY to refine her design direction and bring clarity to her creative process.

Jenny asked CLAY to create a moodboard for an athleisure casual look suitable for a resort setting (Figure~\ref{fig:CIF}-1). CLAY provided several detailed style suggestions (Figure~\ref{fig:CIF}-2), and among them, Jenny was particularly intrigued by the keyword "Summer Breeze Athleisure." She explored how she could further develop the different fashion elements associated with this style. Jenny selected keywords such as "pastel tones of blue, pink, and white, soft colors perfect for summer, and loose, comfortable fits" that evoke a breezy resort atmosphere. To enhance the athleisure vibe, she also chose keywords such as "breathable mesh or lightweight cotton, linen, or stretchy fabrics" (Figure~\ref{fig:CIF}-3). Feeling that the initial moodboard was somewhat a bit plain (Figure~\ref{fig:CIF}-4), Jenny decided to add more flair by incorporating elements of the "Tropical Print Athleisure" style, specifically "printed patterns and thin strap details," to rework the moodboard (Figure~\ref{fig:CIF}-5). Ultimately, the final moodboard, with its bright colors, tropical patterns, and fitted silhouettes, provided her with significant inspiration for developing her resort-inspired athleisure design (Figure~\ref{fig:CIF}-6).

As Jenny expanded her design based on the moodboard, she felt inspired to incorporate the bra top, pants, and tropical pattern into her design. She entered these images into CLAY (Figure~\ref{fig:CIF}-7), and when she reviewed the suggested usage keywords (Figure~\ref{fig:CIF}-8), she was surprised to find more elements she could use than she had originally imagined. Jenny extracted and refined keywords such as "stripe detail under the bust, light pastel blue and white, drawstring waist, full print pattern, and thin straps characteristic of the bra top style." After customizing the prompt with these elements, she generated new images (Figure~\ref{fig:CIF}-9). After reviewing the results, Jenny realized that there were too many pants in her current design (Figure~\ref{fig:CIF}-10), which was different from the vacation vibe she was going for. She thought that adding an active skirt might be more in line with her design direction. She then added the keyword "active skirt" (Figure~\ref{fig:CIF}-11), and CLAY suggested a tennis skirt silhouette that fit her vision perfectly (Figure~\ref{fig:CIF}-12). Jenny felt that this design choice resonated well with her concept and submitted the final design for the competition.

\section{Study Design}
Our user study aims to evaluate CLAY and gain insights for the design of an AI-based interactive system to support balancing vagueness. We recruited 12 participants, including fashion professionals, for a 75-minute study involving two rounds of moodboard creation and design tasks. The study simulates the early, vague stages of fashion design, with participants starting from a vague style concept rather than replicating existing designs. Through surveys, the Creativity Support Index, NASA-TLX questionnaires, and semi-structured interviews, we assessed the system's usability, efficiency, and impact on the creative process.

\subsection{Participants}
We recruited participants currently working as fashion professionals through an online community, following the same recruitment community as used in the formative study. Note that these participants were newly recruited for the CLAY evaluation study and were different from those in the formative study. A total of 12 participants were recruited, excluding two applicants who did not meet the task requirements. The final participants included three males and seven females, while fashion design experience ranged from 2 to 5 years. Each participant was invited to a university laboratory for the user study. Our study was approved by the Institutional Review Board, and participant consent was obtained prior to the study. Each participant was compensated \$50 for their participation.

\subsection{Study procedure}
Our study lasted an average of 75 minutes, during which participants completed two rounds of moodboard creation and design tasks in two different environments: CLAY and the baseline (DALL-E). Rather than replicating existing designs, participants were asked to start with a style and create a new moodboard and design. This simulated the early vague state that fashion designers typically experience, as opposed to working from a clearly defined design idea. To prevent participants from deviating from the goal of balancing vagueness, the following style keywords were randomly provided in each tasks: feminine, vintage, sporty, chic, hip-hop, and futuristic. These keywords were selected through informal discussions with fashion designers to ensure that each style was distinct and non-overlapping. Based on the given style, participants could freely input their first prompt. During interviews, two participants mentioned that they had based their work on previous designs and were excluded from the final evaluation. The order of tool usage was counterbalanced across participants.

Each round began with a 2-minute tutorial on the given system, during which participants were familiarized with the tool using sample images. Participants then created their moodboards, expanded their ideas based on the moodboards, and sketched their designs on paper using a pen. After each round, participants completed a post-task survey. A 10-minute break was provided between the two rounds. After completing both rounds, a 20-minute semi-structured interview was conducted to discuss the differences between the two conditions and the impact of the tools on the ideation process.

The baseline system had a similar interface to CLAY, but it did not include key features of CLAY, such as keyword extraction, hierarchical recommendations, and combination image generation. Instead, the baseline system was designed to allow users to start with an initial vague prompt and freely enter prompts throughout the design process. To evaluate the efficiency of CLAY's features in design  process, the focus was placed on evaluating CLAY's feature design rather than the impact of the AI functionality. To prevent model performance from affecting the study's outcomes, the DALL-E API integrated into CLAY was also used in the baseline system.

\subsection{Data collection}

The survey after each round included questions about the usefulness of the given system for creating moodboards and developing designs from vagueness. For the quantitative evaluation, we conducted a web-based survey. In this survey, we measured the usability of the system using commonly applied user experience evaluation criteria: effectiveness, productivity, usefulness, control over activities, ease of accomplishment, time savings, ability to meet needs, and expected performance~\citep{lund2001measuring}. In addition, we included five questions to assess participants' self-perceived experience with the AI system~\cite{wu2022ai}. Participants rated each task on a 7-point Likert scale, ranging from (1) strongly disagree to (7) strongly agree. The survey also included the Creativity Support Index~\cite{cherry2014quantifying} and the NASA-TLX questionnaire~\cite{hart2006nasa}. We also collected the number of interactions participants made and analyzed how efficiently they could reach the final design with fewer interactions.

After both rounds, we conducted a 15-minute semi-structured interview to explore the differences between the two conditions and the impact of the tools on creating moodboards and designs from vagueness. To facilitate participants' recall of their task performance, we showed them the screens they interacted with while using CLAY. All interviews were audio-recorded. We asked participants to share their thoughts and experiences on the following topics: (1) the main differences in completing the moodboard and design process using the baseline and CLAY, (2) how they accomplished the task using both the baseline and CLAY, (3) which components were helpful or unhelpful in exploring from the initial concept and achieving the intended images using the baseline and CLAY, and (4) how using the baseline and CLAY could support their actual design process.

\subsection{Analysis methods}
In the quantitative analysis of survey responses, we sought to examine whether there was a significant difference in the perception between CLAY and the baseline. We analyzed the differences in survey responses and the number of interactions between the two groups using a $t$-test (normality check was performed).

In the qualitative analysis, we aimed to gain a deeper understanding of the usability perception and the difference in system usage between the two system conditions. The two researchers (authors of this paper) reviewed the screen and voice recordings and wrote down the participants' comments in detail. Next, we conducted an inductive thematic analysis~\citep{vaismoradi2013content}. The two researchers conducted open coding to compare the differences between baseline and CLAY. We conducted three two-hour meetings and conducted axial coding. We resolved conflicts between the coders' results through iterative discussions and comparisons of the coded results.

\section{Results}
\subsection{Moodboard and design creation support capabilities (Quantitative)}

\subsubsection{System usability and perceived experience}

Participants rated CLAY as more useful than the baseline in the process of creating moodboards and finalizing designs. As shown in Table~\ref{tab:ttest1}, CLAY significantly increased the effectiveness and productivity of the design process and provided significant benefits to participants. Specifically, CLAY allowed participants to better control their design tasks, which helped them complete their work more easily and save time. Additionally, the system presented participants with images that met their design needs and delivered results that aligned with their expectations.

\begin{table*}[]
\caption{$t$-test results for the user experience evaluation, self-perceived experience with the AI, and the number of interactions between CLAY and the Baseline system (+p < 0.1, *p < 0.05, **p < 0.01, ***p < 0.001).}
\Description{This table presents t-test results comparing user experience evaluation, self-perceived experience with the AI, and interaction counts between the CLAY system and a baseline system. The table includes the mean, standard deviation (std), p-value, and significance (Sig) levels for various user experience categories. Key results indicate that the CLAY system significantly outperformed the baseline in terms of effectiveness (p = 0.006, **), productivity (p = 0.041, *), usefulness (p = 0.012, *), controlling activities (p < 0.001, ***), accomplishing tasks more easily (p = 0.005, **), saving time (p = 0.037, *), meeting user needs (p = 0.003, **), and matching goals (p = 0.006, **). Other significant results include self-perception aspects such as thinking through tasks (p = 0.006, **), controllability (p = 0.048, *), and collaboration (p = 0.002, **). The number of interactions was higher for the baseline system, but the difference was statistically significant (p = 0.049, *).}
\resizebox{0.8\textwidth}{!}{
\begin{tabular}{rrcccccc}
\hlineB{2}
\hline
\multicolumn{1}{l}{}                                                                                                                                          & \multicolumn{1}{l}{}                        & \multicolumn{2}{c}{\textbf{CLAY}}                     & \multicolumn{2}{c}{\textbf{Baseline}}                   & \multicolumn{2}{c}{\textbf{Statistics}}              \\ \cline{3-8} 
\multicolumn{1}{l}{}                                                                                                                                          & \multicolumn{1}{l}{}                        & \textbf{mean}            & \textbf{std}               & \textbf{mean}              & \textbf{std}               & \textbf{$p$}                            & \textbf{Sig} \\ \hline
\multicolumn{2}{r|}{\textbf{Effective}}                                                                                                                                                                     & \multicolumn{1}{c|}{6.5} & \multicolumn{1}{c|}{0.972} & \multicolumn{1}{c|}{1.6} & \multicolumn{1}{c|}{4.6}   & \multicolumn{1}{c|}{0.006}            & **           \\ \hline
\multicolumn{2}{r|}{\textbf{Productive}}                                                                                                                                                                    & \multicolumn{1}{c|}{6.3} & \multicolumn{1}{c|}{1.252} & \multicolumn{1}{c|}{4.8}   & \multicolumn{1}{c|}{1.751} & \multicolumn{1}{c|}{0.041}            & *            \\ \hline
\multicolumn{2}{r|}{\textbf{Useful}}                                                                                                                                                                        & \multicolumn{1}{c|}{6.4} & \multicolumn{1}{c|}{0.843} & \multicolumn{1}{c|}{5.0}     & \multicolumn{1}{c|}{1.333} & \multicolumn{1}{c|}{0.012}            & *            \\ \hline
\multicolumn{2}{r|}{\textbf{Control Activities}}                                                                                                                                                            & \multicolumn{1}{c|}{6.3} & \multicolumn{1}{c|}{0.823} & \multicolumn{1}{c|}{3.9}   & \multicolumn{1}{c|}{1.663} & \multicolumn{1}{c|}{\textless{}0.001} & ***          \\ \hline
\multicolumn{2}{r|}{\textbf{Accomplish Easier}}                                                                                                                                                             & \multicolumn{1}{c|}{6.5} & \multicolumn{1}{c|}{0.707} & \multicolumn{1}{c|}{4.4}   & \multicolumn{1}{c|}{1.955} & \multicolumn{1}{c|}{0.005}            & **           \\ \hline
\multicolumn{2}{r|}{\textbf{Save Time}}                                                                                                                                                                     & \multicolumn{1}{c|}{6.6} & \multicolumn{1}{c|}{1.265} & \multicolumn{1}{c|}{4.9}   & \multicolumn{1}{c|}{2.025} & \multicolumn{1}{c|}{0.037}            & *            \\ \hline
\multicolumn{2}{r|}{\textbf{Meet Needs}}                                                                                                                                                                    & \multicolumn{1}{c|}{6.2} & \multicolumn{1}{c|}{0.919} & \multicolumn{1}{c|}{4.2}   & \multicolumn{1}{c|}{1.619} & \multicolumn{1}{c|}{0.003}            & **           \\ \hline
\multicolumn{2}{r|}{\textbf{De Expected}}                                                                                                                                                                   & \multicolumn{1}{c|}{4.9} & \multicolumn{1}{c|}{1.663} & \multicolumn{1}{c|}{3.2}   & \multicolumn{1}{c|}{1.476} & \multicolumn{1}{c|}{0.026}            & *            \\ \hline
\multicolumn{1}{r|}{\multirow{5}{*}{\textbf{\begin{tabular}[c]{@{}r@{}}participants'\\ self-perceived \\ experience \\ using the AI \\ system\end{tabular}}}} & \multicolumn{1}{r|}{\textbf{Match Goal}}    & \multicolumn{1}{c|}{6.2} & \multicolumn{1}{c|}{1.135} & \multicolumn{1}{c|}{4.2}   & \multicolumn{1}{c|}{1.687} & \multicolumn{1}{c|}{0.006}            & **           \\ \cline{2-8} 
\multicolumn{1}{r|}{}                                                                                                                                         & \multicolumn{1}{r|}{\textbf{Think Through}} & \multicolumn{1}{c|}{6.6} & \multicolumn{1}{c|}{0.669} & \multicolumn{1}{c|}{4.4}   & \multicolumn{1}{c|}{2.119} & \multicolumn{1}{c|}{0.006}            & **           \\ \cline{2-8} 
\multicolumn{1}{r|}{}                                                                                                                                         & \multicolumn{1}{r|}{\textbf{Transparent}}   & \multicolumn{1}{c|}{6.2} & \multicolumn{1}{c|}{0.919} & \multicolumn{1}{c|}{5.3}   & \multicolumn{1}{c|}{1.337} & \multicolumn{1}{c|}{0.096}            & \textbf{+}   \\ \cline{2-8} 
\multicolumn{1}{r|}{}                                                                                                                                         & \multicolumn{1}{r|}{\textbf{Controllable}}  & \multicolumn{1}{c|}{6.3} & \multicolumn{1}{c|}{1.16}  & \multicolumn{1}{c|}{4.9}   & \multicolumn{1}{c|}{1.729} & \multicolumn{1}{c|}{0.048}            & *            \\ \cline{2-8} 
\multicolumn{1}{r|}{}                                                                                                                                         & \multicolumn{1}{r|}{\textbf{Collaborative}} & \multicolumn{1}{c|}{6.5} & \multicolumn{1}{c|}{1.08}  & \multicolumn{1}{c|}{4.2}   & \multicolumn{1}{c|}{1.619} & \multicolumn{1}{c|}{0.002}            & **           \\ \hline
\multicolumn{2}{r|}{\textbf{Interaction Count}}                                                                                                                                                             & \multicolumn{1}{c|}{6.6} & \multicolumn{1}{c|}{2.17}  & \multicolumn{1}{c|}{11.4}  & \multicolumn{1}{c|}{6.87}  & \multicolumn{1}{c|}{0.049}            & *            \\ \hline
\hlineB{2}
\end{tabular}
}
\label{tab:ttest1}

\end{table*}

The perceived experience of using the AI-based system, as reflected in the survey results, provided more concrete evidence of this benefits. As shown in Table~\ref{tab:ttest1}, CLAY received significantly higher ratings in areas such as match goal, think through, controlllable, and collaborative, with a slightly significant difference in transparency. This indicates that participants felt they could collaborate and control the system more easily with CLAY. As a result, participants were able to reflect more deeply on their design outcomes, leading to more satisfying final results.

Notably, this benefit was achieved with relatively fewer interactions. According to Table~\ref{tab:ttest1}, the participants completed their designs with fewer interactions using CLAY compared to the baseline. This demonstrates CLAY’s ability to support an efficient design process with minimal prompt iterations. In interviews, participants reported that while they had to constantly adjust prompts to generate desired images with the baseline, they were able to refine images in the desired direction with fewer interactions when using CLAY.

\begin{table*}[]
\caption{$t$-test results for NASA-TLX total score and subcomponents between CLAY and the Baseline system (+p < 0.1, *p < 0.05, **p < 0.01).}
\Description{This table presents the t-test results for NASA-TLX total score and subcomponents comparing the CLAY system and the Baseline system. The table includes mean, standard deviation (std), p-value, and significance (Sig) levels for different categories. The total NASA-TLX score shows that CLAY (mean = 29, std = 2.17) significantly outperforms the Baseline (mean = 46.6, std = 6.87) with a p-value of 0.04 (*). For the subcomponents, CLAY shows lower scores in Mental (mean = 21) and Physical (mean = 11) compared to the Baseline, but these differences are not statistically significant. Temporal shows a noticeable difference, though not significant (p = 0.110). Effort shows a marginally significant result in favor of CLAY (p = 0.087, +). Performance is significantly higher for CLAY (mean = 83, p = 0.002, **), while Frustration is lower for CLAY, but the difference is not statistically significant (p = 0.152).}
\resizebox{0.72\textwidth}{!}{
\begin{tabular}{rrcccccc}
\hlineB{2}
\hline
\multicolumn{1}{l}{}                                    & \multicolumn{1}{l}{}                      & \multicolumn{2}{c}{\textbf{CLAY}}                       & \multicolumn{2}{c}{\textbf{Baseline}}                  & \multicolumn{2}{c}{\textbf{Statistics}}   \\ \cline{3-8} 
\multicolumn{1}{l}{}                                    & \multicolumn{1}{l}{}                      & \textbf{mean}              & \textbf{std}               & \textbf{mean}             & \textbf{std}               & \textbf{P}                 & \textbf{Sig} \\ \hline
\multicolumn{1}{r|}{\multirow{7}{*}{\textbf{NASA-TLX}}} & \multicolumn{1}{r|}{\textbf{Score}}       & \multicolumn{1}{c|}{29.0}    & \multicolumn{1}{c|}{15.87}  & \multicolumn{1}{c|}{46.6} & \multicolumn{1}{c|}{19.59}  & \multicolumn{1}{c|}{0.04}  & *            \\ \cline{2-8} 
\multicolumn{1}{r|}{}                                   & \multicolumn{1}{r|}{\textbf{Mental}}      & \multicolumn{1}{c|}{21.0}    & \multicolumn{1}{c|}{26.23} & \multicolumn{1}{c|}{37.5} & \multicolumn{1}{c|}{29.37} & \multicolumn{1}{c|}{0.202} &              \\ \cline{2-8} 
\multicolumn{1}{r|}{}                                   & \multicolumn{1}{r|}{\textbf{Physical}}    & \multicolumn{1}{c|}{11.0}    & \multicolumn{1}{c|}{24.47} & \multicolumn{1}{c|}{15.5} & \multicolumn{1}{c|}{21.27} & \multicolumn{1}{c|}{0.666} &              \\ \cline{2-8} 
\multicolumn{1}{r|}{}                                   & \multicolumn{1}{r|}{\textbf{Temporal}}    & \multicolumn{1}{c|}{17.0}    & \multicolumn{1}{c|}{24.06} & \multicolumn{1}{c|}{37.0}   & \multicolumn{1}{c|}{28.98} & \multicolumn{1}{c|}{0.110} & \textbf{}    \\ \cline{2-8} 
\multicolumn{1}{r|}{}                                   & \multicolumn{1}{r|}{\textbf{Effort}}      & \multicolumn{1}{c|}{76.0}    & \multicolumn{1}{c|}{18.07} & \multicolumn{1}{c|}{63.0}   & \multicolumn{1}{c|}{13.78} & \multicolumn{1}{c|}{0.087} & +            \\ \cline{2-8} 
\multicolumn{1}{r|}{}                                   & \multicolumn{1}{r|}{\textbf{Performance}} & \multicolumn{1}{c|}{83.0}    & \multicolumn{1}{c|}{15.13} & \multicolumn{1}{c|}{51.0}   & \multicolumn{1}{c|}{23.78} & \multicolumn{1}{c|}{0.002} & **           \\ \cline{2-8} 
\multicolumn{1}{r|}{}                                   & \multicolumn{1}{r|}{\textbf{Frustration}} & \multicolumn{1}{c|}{18.5} & \multicolumn{1}{c|}{19.44} & \multicolumn{1}{c|}{36.5} & \multicolumn{1}{c|}{32.75} & \multicolumn{1}{c|}{0.152} &              \\ \hline
\hlineB{2}
\end{tabular}
}
\label{tab:ttest2}
\end{table*}

\subsubsection{Perceived workload}

In the overall workload evaluation conducted using NASA-TLX, CLAY scored statistically significantly lower than the baseline (Table~\ref{tab:ttest2}). To examine the subcomponents of NASA-TLX where CLAY was rated as providing better support than the baseline, individual $t$-tests were performed on each of the six dimensions. The analysis revealed a significant difference in the Performance dimension and a marginally significant difference in the Effort dimension, indicating that CLAY improved the efficiency of task performance. However, no significant differences were found between the two systems on the Mental, Physical, Temporal, and Frustration dimensions. This suggests that both CLAY and the baseline, by using a prompt-based image generation approach, did not cause users to experience excessive burden during task performance.

\begin{table*}[]
\caption{$t$-test results for CSI total score and subcomponents between CLAY and the Baseline system (+p < 0.1, *p < 0.05, **p < 0.01).}
\Description{This table presents the t-test results for the Creativity Support Index (CSI) total score and subcomponents comparing the CLAY system and the Baseline system. The table includes the mean, standard deviation (std), p-value, and significance (Sig) levels for different categories. The total CSI score shows that CLAY (mean = 84.6, std = 13.47) significantly outperforms the Baseline (mean = 62.9, std = 22.7) with a p-value of 0.018 (*). For subcomponents, CLAY shows significant improvement in Enjoyment (p = 0.008, **), while Exploration shows marginal significance (p = 0.078, +). Results Worth Effort also shows a significant result in favor of CLAY (p = 0.035, *). Other subcomponents, such as Expressiveness, Immersion, and Collaboration, show higher means for CLAY but without significant differences.}
\resizebox{0.74\textwidth}{!}{
\begin{tabular}{rrcccccc}
\hlineB{2}
\hline
\multicolumn{1}{l}{}                                                                                                  & \multicolumn{1}{l}{}                               & \multicolumn{2}{c}{\textbf{CLAY}}                          & \multicolumn{2}{c}{\textbf{Baseline}}                  & \multicolumn{2}{c}{\textbf{Statistics}}   \\ \cline{3-8} 
                                                                                                                      &                                                    & \textbf{mean}             & \textbf{std}                   & \textbf{mean}             & \textbf{std}               & \textbf{P}                 & \textbf{Sig} \\ \hline
\multicolumn{1}{r|}{\multirow{7}{*}{\textbf{\begin{tabular}[c]{@{}r@{}}Creativity \\ Support \\ Index\end{tabular}}}} & \multicolumn{1}{r|}{\textbf{Score}}                & \multicolumn{1}{c|}{84.6} & \multicolumn{1}{c|}{13.47} & \multicolumn{1}{c|}{62.9} & \multicolumn{1}{c|}{22.70}  & \multicolumn{1}{c|}{0.018} & *            \\ \cline{2-8} 
\multicolumn{1}{r|}{}                                                                                                 & \multicolumn{1}{r|}{\textbf{Enjoyment}}            & \multicolumn{1}{c|}{88.0}   & \multicolumn{1}{c|}{12.52} & \multicolumn{1}{c|}{58.0}   & \multicolumn{1}{c|}{29.08} & \multicolumn{1}{c|}{0.008} & **           \\ \cline{2-8} 
\multicolumn{1}{r|}{}                                                                                                 & \multicolumn{1}{r|}{\textbf{Exploration}}          & \multicolumn{1}{c|}{79.0}   & \multicolumn{1}{c|}{16.96}     & \multicolumn{1}{c|}{57.0}   & \multicolumn{1}{c|}{33.10} & \multicolumn{1}{c|}{0.078} & +            \\ \cline{2-8} 
\multicolumn{1}{r|}{}                                                                                                 & \multicolumn{1}{r|}{\textbf{Expressiveness}}       & \multicolumn{1}{c|}{81.0}   & \multicolumn{1}{c|}{18.53}     & \multicolumn{1}{c|}{65.5} & \multicolumn{1}{c|}{24.99} & \multicolumn{1}{c|}{0.113} & \textbf{}    \\ \cline{2-8} 
\multicolumn{1}{r|}{}                                                                                                 & \multicolumn{1}{r|}{\textbf{Immersion}}            & \multicolumn{1}{c|}{84.5} & \multicolumn{1}{c|}{18.17}     & \multicolumn{1}{c|}{68.5} & \multicolumn{1}{c|}{33.75} & \multicolumn{1}{c|}{0.203} &              \\ \cline{2-8} 
\multicolumn{1}{r|}{}                                                                                                 & \multicolumn{1}{r|}{\textbf{Results Worth Effort}} & \multicolumn{1}{c|}{90.5} & \multicolumn{1}{c|}{11.17}     & \multicolumn{1}{c|}{71.5} & \multicolumn{1}{c|}{23.81} & \multicolumn{1}{c|}{0.035} & *            \\ \cline{2-8} 
\multicolumn{1}{r|}{}                                                                                                 & \multicolumn{1}{r|}{\textbf{Collaboration}}        & \multicolumn{1}{c|}{82.0}   & \multicolumn{1}{c|}{16.19}     & \multicolumn{1}{c|}{61.5} & \multicolumn{1}{c|}{34.65} & \multicolumn{1}{c|}{0.107} &              \\ \hline
\hlineB{2}
\end{tabular}
}
\label{tab:ttest3}
\end{table*}

\subsubsection{Creativity Support Index}

CLAY was rated significantly higher than the baseline in terms of creativity support (Table~\ref{tab:ttest3}). To examine the specific components in which CLAY provided better creative support than the baseline, $t$-tests were conducted on each of the six dimensions. Significant differences in favor of CLAY were found only in the Enjoyment and Results Worth Effort dimensions, with a marginally significant difference in the Exploration dimension. Participants reported that they enjoyed creating moodboards and designs with less effort using CLAY, the baseline was found to be helpful when participants were stuck during ideation, stimulating creativity in such situations.

\subsection{Balancing vagueness capabilities (Qualitative)}

\subsubsection{CLAY main structure}

\begin{figure*}[]
\centering
\includegraphics[width=0.9\textwidth]{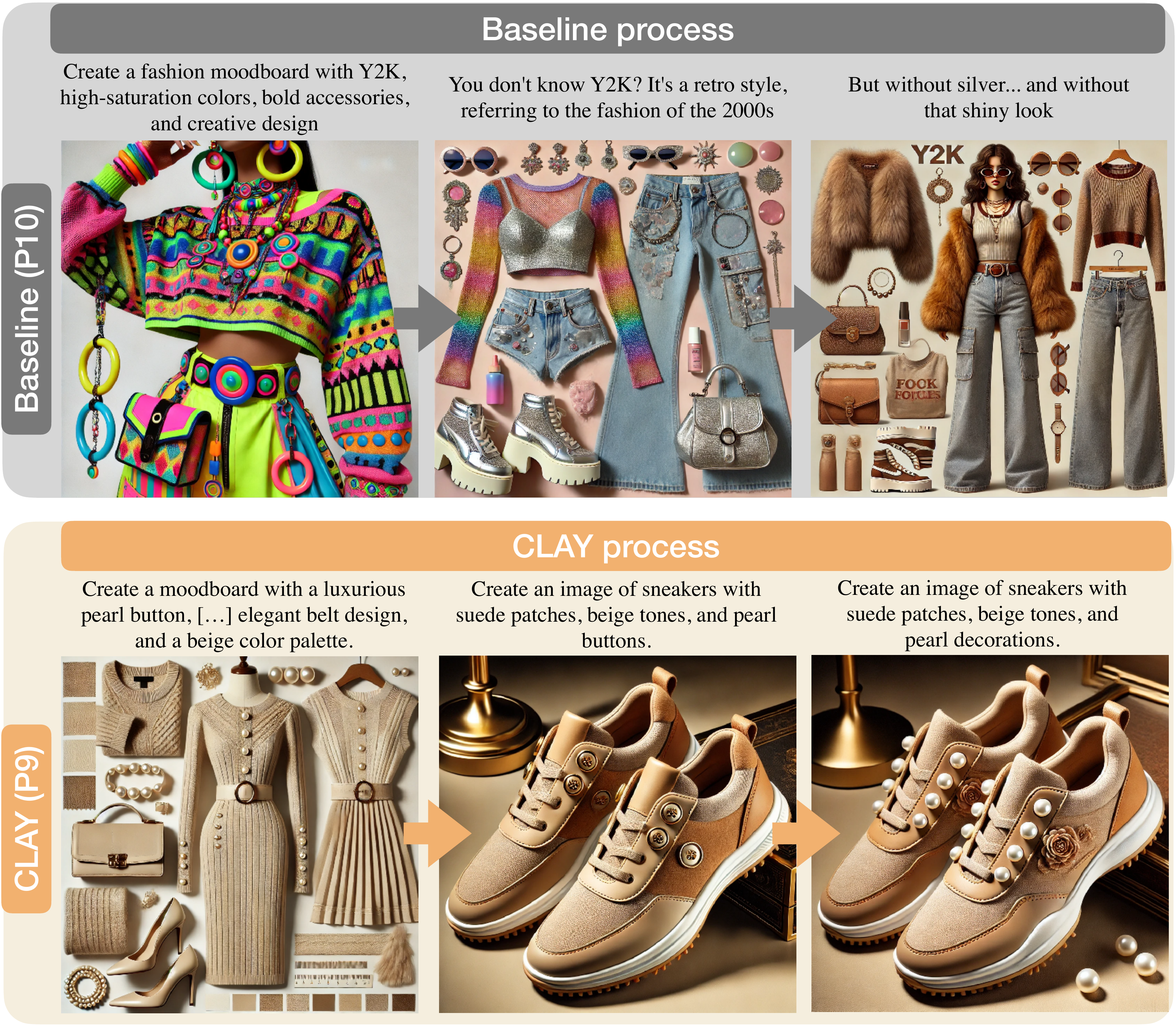}
\vspace{-0.2cm}
\caption{The figure contrasts the process of balancing vagueness between the Baseline (top) and CLAY (bottom). The user in the Baseline condition was driven by the results and did not know how to efficiently communicate with the system. In contrast, the user in the CLAY process showed the clear interactions with the system.}
\Description{This figure contrasts the process of balancing vagueness between the Baseline (top) and CLAY (bottom). In the Baseline process, users created a fashion moodboard with a Y2K theme, featuring high-saturation colors, bold accessories, and creative designs. The process also highlights confusion when specifying style preferences, as the user struggled with vague communication. In contrast, the CLAY process involved creating a moodboard with a luxurious pearl button, beige tones, and elegant design elements. Users were able to refine their ideas, specifying details like sneakers with suede patches, beige tones, and pearl buttons, demonstrating clear and effective interaction with the system.}
\label{fig:Whole_ex}
\end{figure*}

\paragraph{\textbf{Baseline made users feel driven by the results.}} Participants noted that the images changed abruptly with each prompt input with no consistency. They mentioned that this might be helpful when no ideas were forming, but overall made it difficult to develop the designs and led to a failure to balance vagueness. In other words, it was difficult to maintain consistency in the process of gradually refining ideas from an initial state of vagueness and failing to make the design direction concrete. \textit{"When designing, sometimes you get stuck and no ideas come to mind. In those moments, I think it would be useful. [...] But even if the design looks good, it’s important that it organically shows how the design emerged from certain elements of the moodboard, and that didn’t happen, so it wasn’t helpful"} (P7). Participants made several attempts to correct the results but eventually gave up (Figure~\ref{fig:Whole_ex}). \textit{"I wanted to create a new Y2K design, so I asked for a bold, highly saturated, creative Y2K accessory, but the result was too rustic. [...] When I asked for more of a Y2K feel, it became too shiny. [...] Then I asked to remove the shine, and it completely changed to brown tones. [...] I gave up communicating with it halfway through."} (P10).

\paragraph{\textbf{Users guided the results with CLAY}} 
CLAY helped balance vagueness through a step-by-step approach in which users viewed hierarchical text generated from vague prompts, refined their prompts based on that text, and then viewed combination images. Users reported that viewing the text helped them exploit vagueness and then reduce it when generating images based on more specific prompts. In particular, users responded positively to various possibilities they could explore by reviewing the sub-fashion element texts, which could branch out from an initial vague style and mood. \textit{"The moodboard shouldn't be too generic [...] It's important to be distinctive. I found it really helpful to see how the keywords could develop from the initial simple ones and use them to create a better moodboard"} (P8). Additionally, users found the fashion element texts suggested for the moodboard useful in fleshing out their designs (Figure~\ref{fig:Whole_ex}). \textit{"I wanted to design sneakers based on the moodboard. I liked the keywords beige, knit, and pearl [...] the sneakers I wanted came out perfectly! [...] Having multiple keywords for details allowed me to create more specific results, which I think led to a better outcome"} (P9).

\paragraph{\textbf{CLAY shortens the process.}} Participants evaluated that CLAY's step-by-step approach allowed them to complete designs and moodboards with fewer interactions compared to the baseline. \textit{"With the baseline, I had to go through more interactions [...] but with CLAY, I could create a moodboard and design quickly with fewer interactions!"} (P4). Participants also reported that CLAY's efficient approach significantly shortened the image search process and helped them generate many designs in a short amount of time. \textit{"There are times when I have to search for numerous images to create a moodboard and then generate multiple designs in an unreasonable amount of time, but CLAY does it quickly and exactly the way I want it to"} (P1).

\subsubsection{CLAY for moodboard stage}

\begin{figure*}[]
\centering
\includegraphics[width=\textwidth]{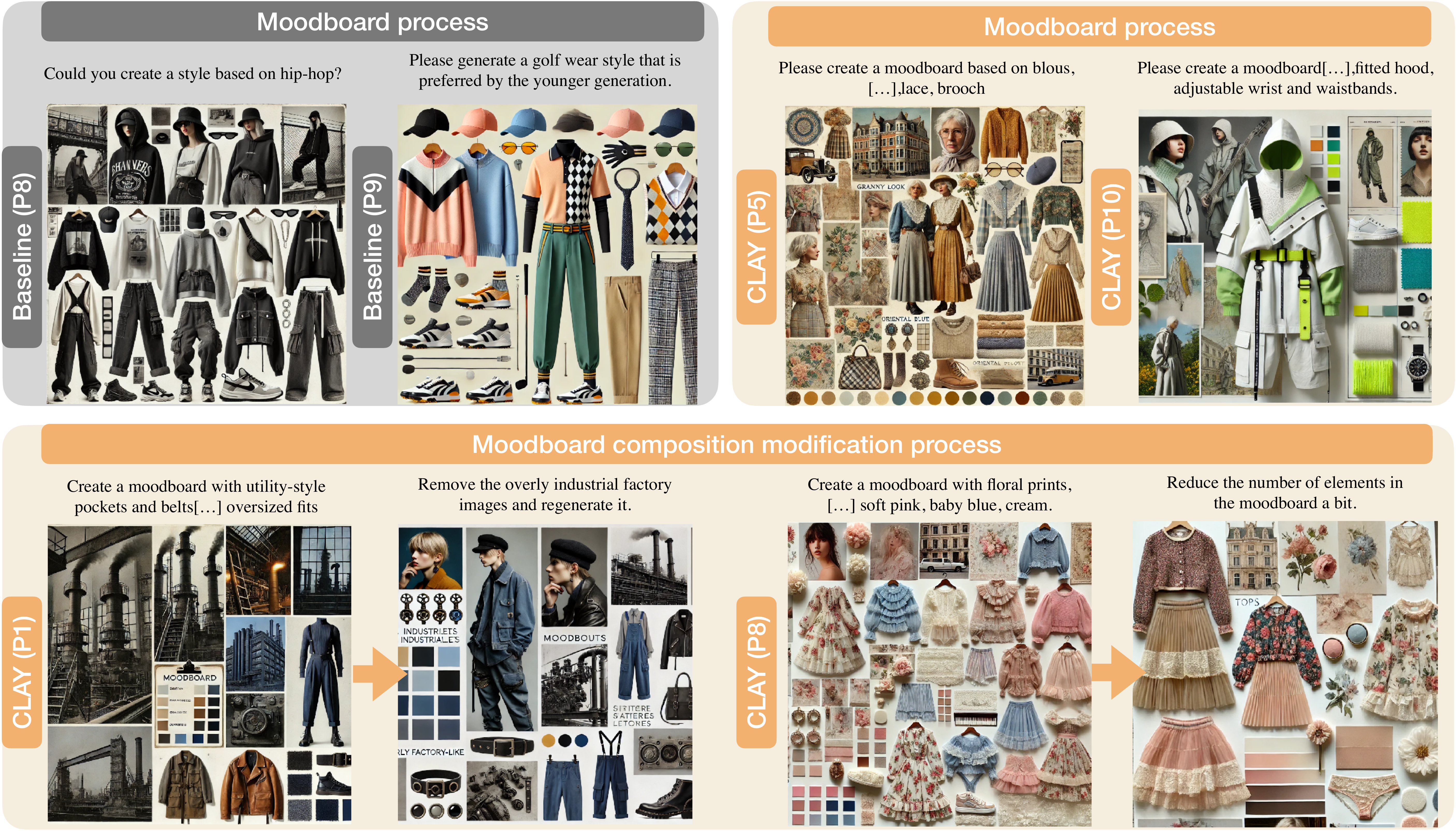}
\vspace{-0.2cm}
\caption{\textbf{(Moodboard stage)} The results of the balancing vagueness process with the moodboard. The top-left row shows moodboards created using the Baseline system, while the top-right row shows moodboards created using CLAY. The bottom row shows examples of how users modified the moodboard composition more proactively with CLAY.}
\Description{This figure illustrates the results of the balancing vagueness process in the "Moodboard stage." The top-left row shows moodboards created using the Baseline system, where users generated styles based on prompts like hip-hop or golf wear for younger generations. The top-right row displays moodboards created using the CLAY system, including detailed prompts such as creating a moodboard based on blouses, lace, and brooches. The bottom row demonstrates how users proactively modified the moodboard compositions with CLAY, including examples like removing industrial factory images or reducing the number of elements in a floral print moodboard.}
\label{fig:Moodboard_ex}
\end{figure*}

\paragraph{\textbf{The generalized results of the Baseline lacked refinement in style.}} Participants responded that the baseline lacked steps to help them expand their thinking from the initial style, making it difficult to explore diversity from vagueness. "It makes similar images. This doesn't lead to good designs" (P8). P9 highlighted the lack of detail, "It only gives generalized, representative images for the keywords I entered, which makes it a bit challenging to use." The examples of images that P8 and P9 interacted with can be seen in Figure~\ref{fig:Moodboard_ex}-Baseline-Moodboard process part.

\paragraph{\textbf{CLAY supports the exploration of moodboards through hierarchical suggestions.}} CLAY allowed the expansion and exploration of ideas from the initial vague style keywords by presenting various sub-styles and sub-fashion elements in detail, allowing users to explore multiple possibilities. \textit{"I was only vaguely thinking about styles. This system helped me expand my thoughts by presenting different categories of sub-styles, like vintage granny or romantic vintage"} (P5). Another participant mentioned that the detailed sub-fashion elements within the sub-styles allowed them to think outside the box. \textit{"When I thought of outdoor, I initially only imagined typical hiking gear. But from this tool, I realized there could be variations like drapery. [...] I wrote drapery, and was able to develop my ideas"} (P10). The examples of images that P5 and P10 interacted with can be seen in Figure~\ref{fig:Moodboard_ex}-CLAY-Moodboard process part.
 
\paragraph{\textbf{CLAY led users to modify moodboard compositions.}} After generating a moodboard with detailed keyword adjustments, some users modified the moodboard’s composition. \textit{"It seemed like there were too many images, so I asked to reduce it. [...] it looked much better"} (P8). Meanwhile, P1 wanted to increase the number of fashion images. \textit{"I requested a rough, industrial style, but there were more object images, so I asked to add more fashion images. The updated results were very satisfying."}  The examples of images that P1 and P8 interacted with can be seen in Figure~\ref{fig:Moodboard_ex}-CLAY-Moodboard composition modification process part.

\subsubsection{CLAY for design stage}

\begin{figure*}[]
\centering
\includegraphics[width=0.9\textwidth]{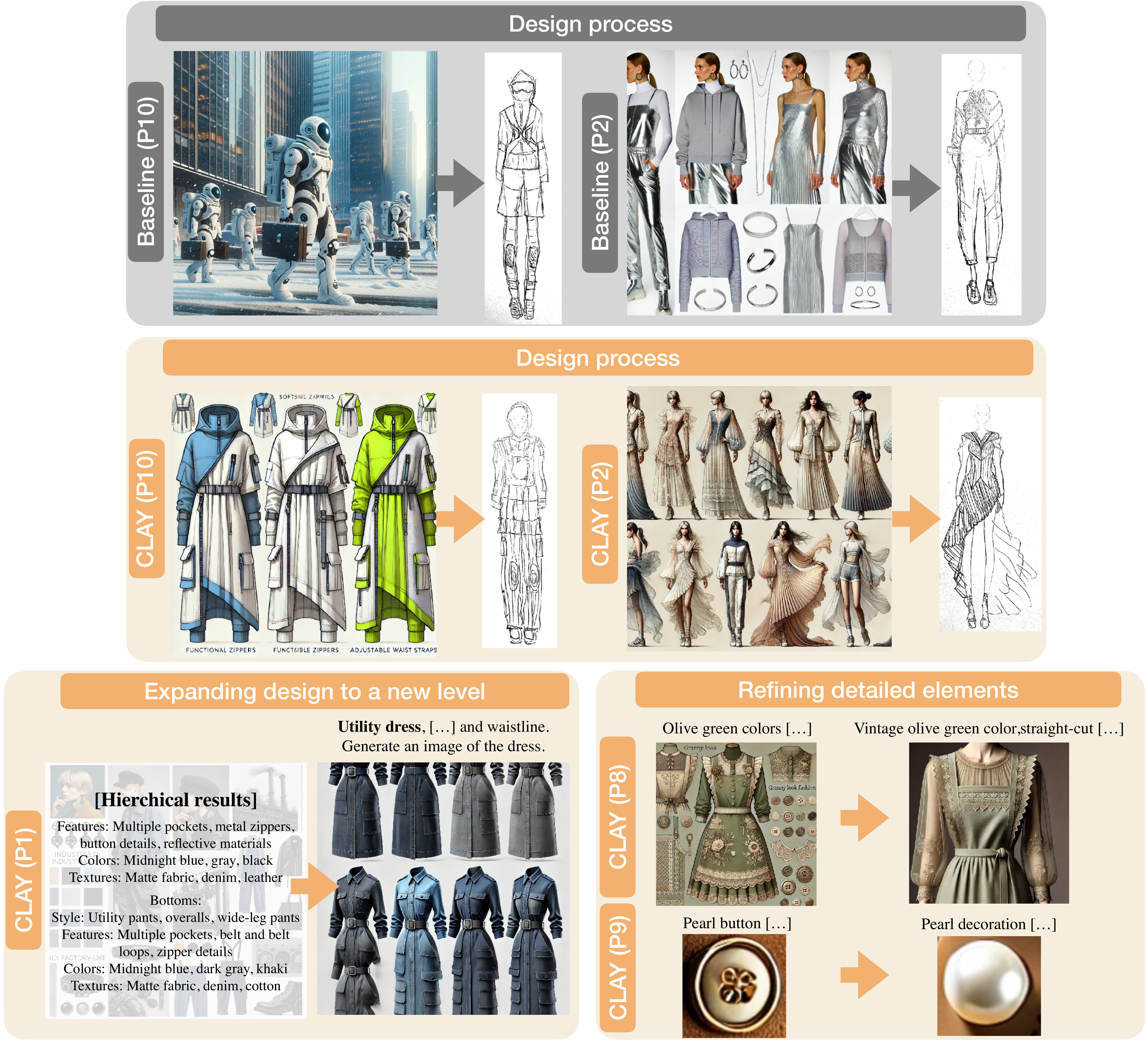}
\vspace{-0.2cm}
\caption{\textbf{(Design stage)} Illustration of the design processes using (top) the Baseline system and (middle) CLAY. At the top, the users minimally referenced the Baseline system in their final design sketches. In contrast, the middle section shows that users referenced CLAY more for their final design sketches. In the bottom row, the left image illustrates a case where the user added new keywords in addition to hierarchical suggestions, while the right image shows the user refining and specifying keywords to better match their intended design.}
\Description{This figure illustrates the "Design stage" of the process, comparing the Baseline system (top) and CLAY (middle). At the top, users in the Baseline system made minimal references to system-generated results in their final design sketches. In contrast, the middle section shows that CLAY users incorporated more specific references for their final sketches, utilizing detailed prompts like utility dresses or olive green colors. The bottom row highlights two examples from CLAY: on the left, users expanded the design by adding new keywords along with hierarchical suggestions, while on the right, users refined detailed elements, such as pearl buttons and decorations, to better match their intended design.}
\label{fig:Design_ex}
\end{figure*}

\paragraph{\textbf{Baseline was only partially referenced for design.}} Participants mentioned that they completed their final sketches by incorporating only a few elements from the designs in the Baseline. \textit{"The AI definitely didn't create what I really wanted. [...] I only referenced the goggles and knee pads, and the rest I just drew from what was in my head."} (P10). Another participant gave up trying to complete their sketch using the Baseline. \textit{"I just gave up because the system did not produce anything useful, so I ended up designing on my own"} (P2). The examples of images that P10 and P2 interacted and sketched can be seen in Figure~\ref{fig:Design_ex}-Baseline-Design process part.

\paragraph{\textbf{CLAY was used more extensively for design.}} In contrast, participants reported that they were able to incorporate more elements into their designs when using CLAY compared to the baseline. \textit{"I referenced it much more during my design process. [...] I changed the pleated details in the image to a padded pleated skirt, and since the fabric looked thin, I thought it would be nice if the padding was also thin"} (P10). Another participant directly incorporated the provided image into the design. \textit{"I really liked this clean, romantic sporty look, so I just used it directly in my sketch."} The examples of images that P10 and P2 interacted and sketched can be seen in Figure~\ref{fig:Design_ex}-CLAY-Design process part.

While participants acknowledged that CLAY provided sufficiently complete images to support the design process from moodboard to sketch, they noted a lack of aesthetic sensitivity. \textit{"The images themselves don't seem aesthetic. [...] But at this stage, what's important is how you explore combine the elements for the design, so the lack of aesthetic refinement is not really an issue"} (P9). In particular, participants felt that the tool lacked the sensitivity required for final design implementation, even though CLAY's images were inspiring. \textit{"The images from the tool will definitely provide inspiration. [...] But when creating clothes from a sketch, I think I would use a tool like Pinterest"} (P6).

\paragraph{\textbf{CLAY expands the design direction to a new level.}} Participants not only use the keyword suggested by the system, but also added new keyword to expand the direction of their designs. P1 came up with the idea of adding contrasting keywords. \textit{"Looking at the element descriptions, there were a lot of masculine keywords, so I was curious about what would happen if I added feminine elements instead. [...] It gave me new ideas, and the design result was better"} (P1). P2 added keywords to steer the design in a more neutral direction. \textit{"The descriptions here felt too feminine and even a bit medieval, so I added sporty keywords to balance it out, and it reflected that really well."} The examples of images that P1 interacted with can be seen in Figure~\ref{fig:Design_ex}-CLAY-Expanding design to a new level part.

\paragraph{\textbf{Refining Detailed Elements with CLAY}} Participants expressed overall satisfaction with the system's results. Some participants further modified the provided results using more specific words to guide the design in the direction they wanted. \textit{"Olive green was fine, but I though it more lighter vintage olive green would be better and changed prompt. [...] Also the skirt was too flashy, so I asked to change it to a more straight-cut design. I liked the refined result more."} (P5). One participant refined prompt using different wording, saying, \textit{"The pearl buttons aren't as pretty as expected. I'll try regenerating them by using 'pearl decorations' instead"} (P9). The examples of images that P8 interacted with can be seen in Figure~\ref{fig:Design_ex}-CLAY-Refining detailed elements.

\section{Discussion}
We propose a novel text-to-image generative AI creativity support tool, CLAY, which guides designers in balancing vagueness. From the user study, we found that users independently embraced and avoided vagueness when using CLAY. Based on our findings, we provide design implications for future text-to-image generative AI tools aimed at balancing vagueness and enhancing creativity.

\subsection{Identifying vagueness explore structure to actively embrace vagueness}

\begin{figure*}[]
\centering
\includegraphics[width=0.9\textwidth]{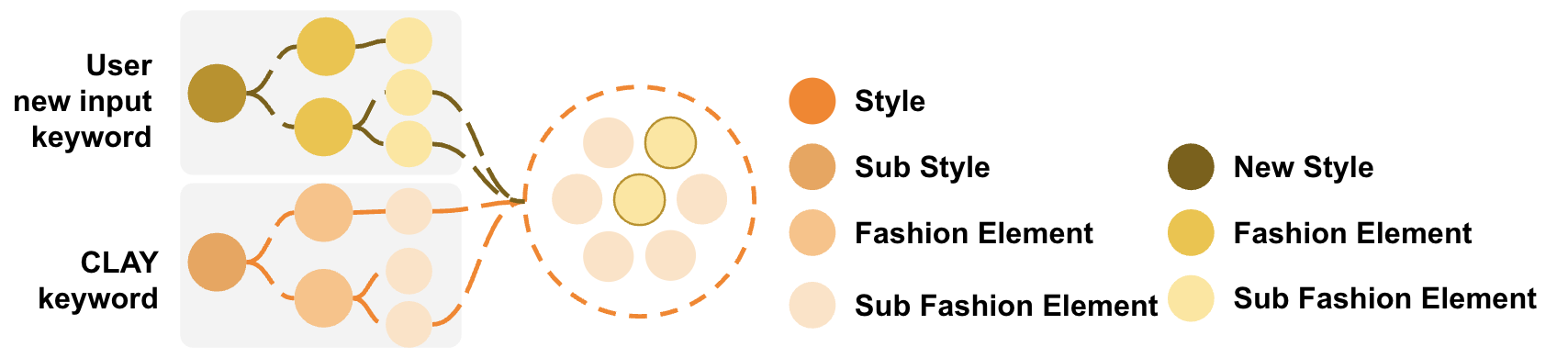}
\vspace{-0.2cm}
\caption{The diagram shows how users add new keywords during the design process, illustrating how they embrace vagueness.}
\Description{This diagram illustrates how users add new keywords during the design process, demonstrating how they handle vagueness. The top section shows new input keywords provided by the user, while the bottom section shows the CLAY system's suggested keywords. The keywords are organized into categories: Style, Sub Style, Fashion Element, and Sub Fashion Element. The diagram highlights how users incorporate and refine these keywords, creating connections between existing suggestions and their own inputs, ultimately expanding and specifying the design's direction.}
\label{fig:Dis1}
\end{figure*}

Participants tended to embrace vagueness more actively by exploring the hierarchical results provided by text-to-image generative AI from a vague prompt. Specifically, designers enjoyed exploring vagueness through more detailed sub-styles and sub-fashion elements starting from the initial concept. Furthermore, when exploring vagueness from moodboard to design, designers embraced vagueness more by not only considering the hierarchical results from CLAY, but also incorporating new style hierarchical elements that were not suggested by CLAY (Figure~\ref{fig:Dis1}). They mentioned that after seeing the hierarchical results, they thought it would be good to add something new as well. This process mirrors the divergent phase in the design process, where ideas are expanded~\cite{eysenck2003creativity}. The key aspect in divergent is not just providing keyword diversity, but encouraging the embrace of vagueness by exploring hierarchical structure.

\begin{figure*}[]
\centering
\includegraphics[width=0.9\textwidth]{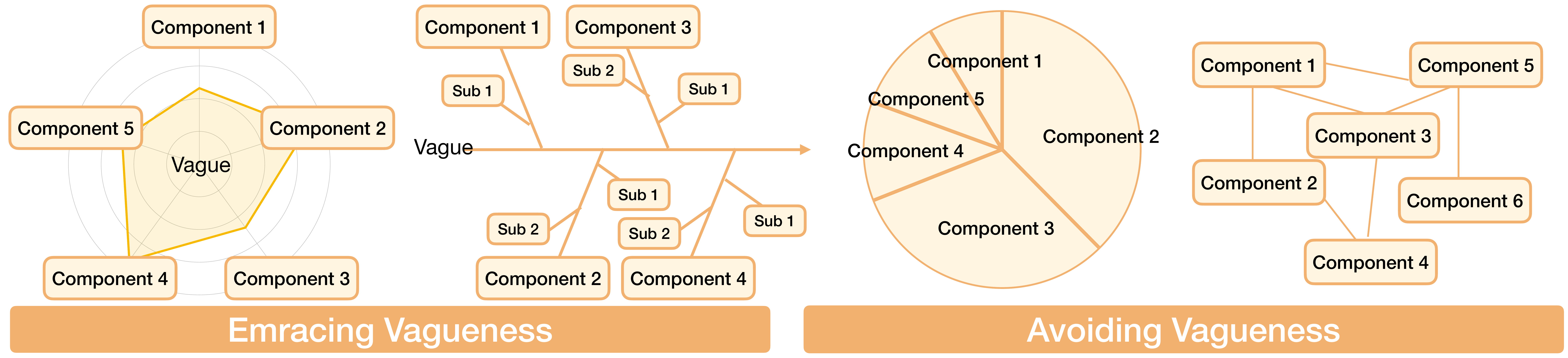}
\caption{Illustrated examples of other domain structures for embracing or avoiding vagueness. Depending on what is identified as the core component that characterizes a domain, different types of structures can be considered to support AI-based interactions.}
\Description{This diagram presents examples of domain structures that either embrace or avoid vagueness. On the left, the "Embracing Vagueness" structure shows a core vague component connected to various sub-components and elements, demonstrating flexibility and multiple interpretations. On the right, the "Avoiding Vagueness" structure organizes components in a more rigid, clearly defined manner, minimizing ambiguity. Depending on the core component that characterizes a domain, different structures can be applied to support AI-based interactions, either allowing for more flexibility or enforcing clearer boundaries.}
\vspace{-0.5cm}
\label{fig:Dis_ex}
\end{figure*}

Based on these findings, we suggest that providing the structure of utilizing vagueness through the text-to-image generative AI is important for promoting the acceptance of vagueness. The structure of vagueness exploration may vary depending on the creative process of each domain. Therefore, by identifying the structure, automatically recognizing the user's workflow stage, and presenting the structure-based results at the right time, domain users can effectively embrace their vagueness. For example, while we propose a hierarchical structure, other types of structures such as a radial structure, where there are differences in the degree of components, or a fishbone structure, which required following a specific order, may be more suitable depending on the domain (Figure~\ref{fig:Dis_ex}). Additionally, to automatically detect the user's workflow stage, identifying domain-specific keywords that represent vagueness could allow the system to recognize the current stage when prompts consist of these vague keywords and provide the appropriate structure-based results.

\subsection{Providing structure-based prompt guidance in appropriate timing to actively avoid vagueness}
Participants showed a tendency to actively avoid vagueness by refining the prompt, combining keywords, and looking at combined results. Specifically, after writing fashion element keywords and looking at the moodboard, some users even changed the component of the moodboard to further avoid vagueness. Additionally, after writing fashion element keywords and looking at CLAY designs, participants tended to further specify their input (e.g., changing "olive green" to "vintage faded olive green") in order to actively reduce vagueness. This is similar to the convergent process, where ideas are refined and focused~\cite{eysenck2003creativity}. CLAY allowed the participants to have their design idea focused, guiding them to specify from the expanded elements of vagueness.

These findings suggest that in order to actively avoid vagueness, it is necessary to provide guidance on the appropriate time to refine prompts based on the structure of reducing vagueness. As noted in the Guidelines for Human-AI Interaction~\cite{amershi2019guidelines}, it is important to interrupt at the appropriate time based on the user's current task and environment, and to provide contextually relevant information. One of the contexts we focused on was the concept of balancing vagueness. Since the structure and timing of prompt refinement can vary across domains,by identifying these, we can effectively guide users in each domain to manage vagueness. For instance, users may want to embrace a higher level of vagueness before deciding to avoid it. Therefore, capturing the right timing could involve detecting moments when users want to focus their ideas, such as when they use prompts such as "Yes, create results based on these keywords," or the system can ask, "Do you want to see results with these keywords?" Additionally, for prompt guidance, if the structure for reducing vagueness involves determining the composition ratio of each component or connecting them with lines (Figure~\ref{fig:Dis_ex}), the system could respond with suggestions such as "Try connecting these elements in your prompt" or "Specify the composition ratio", after detecting the user's intent to focus their ideas.

\subsection{Allowing users to select systems based on vagueness state to facilitate embracement}
Based on the results of our study, we observed that some participants preferred simple information retrieval methods during the absence of ideas in the early stages of the design process. CLAY was found to be helpful in resolving vagueness when there was an initial, vague idea in mind (e.g., style, mood). However, in the first stage, when even the style was undecided, participants tended to prefer simpler methods of gathering information and even enjoyed the randomness and complexity that resulted from simple processes. For instance, they would borrow fashion books from libraries or take pictures in museums to broaden the scope of their thinking. This was supported by both the quantitative and qualitative results, as the Baseline system did not show significant differences in the subcategories of the CSI. Participants also mentioned that the sudden, unexpected results from the Baseline system could be useful when ideas are lacking in the early stages of the design process. Previous studies have also suggested that searching for information with a specific goal may be inefficient in the early stages, and gathering random information may be more effective~\cite{ishikawa2009showing, keller2020uncertainty}.

Based on these findings, we suggest that allowing users to select the system they want based on their current level of idea can further facilitate the acceptance of vagueness. For instance, the system could display random text options that users can click on to view random images. Simultaneously, the system can offer the option of writing prompts below the display to guide the user in balancing vagueness. In the absence of a clear idea, users could explore random text and images, helping them embrace vagueness. However, if users have a fairly clear initial concept in mind, the system could provide guidelines for balancing vagueness. Since the type of information users seek in the absence of a clear idea differs across domains, we suggest further research into the type of information users seek for their task. This information should then be incorporated into the random results provided by the system.

\subsection{The impact of image realism on materialization}
It was confirmed that CLAY effectively supports the process of balancing vagueness, but there was feedback that it lacks the ability to provide stylish and aesthetically refined images. Participants noted that traditional image search tools often lead to unconscious imitation during the design process, while generated images allowed them to design without imitation. Previous research has also confirmed that it is extremely difficult to design in fashion without referencing other images, making it nearly impossible to avoid imitation~\cite{eckert2000sources}. Nonetheless, in the final phase of design, where sketches are translated into actual products, text-to-image generative AI still struggles to fully support this process. For example, tasks such as fabric selection (e.g., brown cowhide vs. brown sheepskin) or adding accessories (e.g., pink gold necklace vs. a white gold necklace) rely heavily on the designer's aesthetic sense. These tasks are aimed at making the designs visually appealing, and developing a designer's sense of "materializing the design" plays a crucial role in achieving the final beauty when the design is produced in real life.

Based on this, in domains where avoiding imitation and focusing on real-world implementation is crucial, it is necessary to guide designers to explore aesthetically detailed images. Rather than allowing users to rely solely on AI-generated images, the system should recognize the appropriate time to move to the materialization phase of the design. A simple way to do this is to ask the user if they would like to see the image in the final stage of vagueness balancing. If the user agrees, the system can automatically present real images similar to the final generated design. This would help users achieve vagueness balancing without imitating existing designs, and provide the necessary aesthetic support for materialization based on their sketches.

\subsection{Limitations and Next Steps}
Our study highlights the importance of balancing vagueness in the design process and offers insights into how text-to-image generative AI should be designed to support balancing vagueness. Nevertheless, there are several limitations that need to be addressed in future research, and there remain open questions that suggest possible next steps.

The insights from the participants in our study may not fully capture the perspectives of all fashion designers regarding balancing vagueness. Nonetheless, the study highlighted that balancing vagueness is an important factor to consider when designing text-to-image generative AI and creativity support tools.

Since our study was limited to 75 minutes, a more longitudinal field study (multiple trials with CLAY) with fashion designers is needed, as the fashion design process typically takes at least 6 months. Such a study would provide a deeper understanding of the impact of vagueness balancing on the design process. Despite these limitations, we believe that our findings contribute to a better understanding of how to consider the level of vagueness when designing a generative AI-based, interactive system.

We focused on the concept of vagueness balancing in order to more effectively support the fashion design process through text-to-image generative AI. However, balancing vagueness is only one of many contextual characteristics in the design process, and different domains may encounter different forms of context. Therefore, we suggest that in future work should identify additional key characteristics within the creative process and explore how these concepts can be integrated into systems to better support creativity.


\section{Conclusion}

This paper focuses on the concept of vagueness in the design process; vagueness can encourage exploration of ideas, while vagueness reduction can prevent directionless design. Based on formative research with fashion design professionals, we identified vagueness as a key property of design and developed CLAY, an interactive system that was designed to help balance vagueness through iterative prompt refinement by integrating the strengths of text-to-image generative AI. We then evaluated CLAY's effectiveness at balancing vagueness with 10 fashion designers. CLAY allowed users to embrace and avoid vagueness with fewer interactions, actively modifying components, refine keywords, and add new terms not provided by the system. These findings highlight key design implications, such as identifying exploration structures and providing timely prompts that help users embrace and avoid vagueness as needed. We hope that our research and suggestions will provide valuable insights for system designers who wish to leverage balancing vagueness in creativity support tools. Additionally, we suggest that consideration of other key components in the design process could further enhance support for the creative workflow support.

\bibliographystyle{ACM-Reference-Format}

\end{document}